\documentclass{JHEP3}
\usepackage{units}
\usepackage{graphicx,epsfig}
\usepackage{amsmath}
\usepackage{enumerate}

\newcommand{\gsim}{\gtrsim}

\newcommand{\Tr}{\mathop\mathrm{Tr}}

\newcommand{\im}{\mathop\mathrm{Im}}
\newcommand{\re}{\mathop\mathrm{Re}}

\newcommand{\be}{\begin{equation}}
\newcommand{\ee}{\end{equation}}
\newcommand{\ba}{\begin{eqnarray}}
\newcommand{\ea}{\end{eqnarray}}

\newcommand{\nuMSM}{\ifmmode\nu\mathrm{MSM}\else$\nu$MSM\fi}

\newcommand{\diag}{\mathop\mathrm{diag}}

\title{Baryon Asymmetry of the Universe in the $\nu$MSM}

\author{Laurent Canetti and Mikhail Shaposhnikov\\
  Institut de Th\'eorie des Ph\'enom\`enes Physiques,\\
  \'Ecole Polytechnique F\'ed\'erale de Lausanne,\\
  CH-1015 Lausanne, Switzerland    \\
  E-mail: \email{Laurent.Canetti@epfl.ch},~~\email{Mikhail.Shaposhnikov@epfl.ch}
}

\abstract {We perform a detailed analysis of baryon asymmetry
generation in the \nuMSM \ (an extension of the Standard Model by
three singlet Majorana fermions with masses below the Fermi scale). 
Fixing a number of  parameters of the \nuMSM \ by the neutrino
oscillation data, we determine the remaining domain of the parameter
space from the requirement of successful baryogenesis. We derive, in
particular, the constraints on the mass splitting of a pair of singlet
fermions, and on the strength of their coupling to ordinary leptons,
essential for  searches of these particles in rare decays of mesons
and in beam-dump experiments with intensive proton beams.}

\begin{document}

\section{Introduction}
\label{se:intro}
There are many ways the baryon asymmetry of the Universe (BAU) could
have arisen. They differ from each other by the nature of baryon and
lepton number non-conservation, CP-violation and by the manner a
departure from thermal equilibrium is realised. For practical purposes
it is important to distinguish between the scenarios which can or
cannot be {\em  experimentally tested}. By experimentally testable we
mean those in which particle physics experiments (limiting the
time-scale by the existing accelerators including the LHC, and by the
ILC, CLIC or an $e\gamma$ collider for the future) could potentially
determine {\em all} parameters of the theory, necessary for
theoretical computation of the {\em amplitude} and of the {\em sign}
of the baryon asymmetry of the Universe\footnote{Of course, the
requirement of testability  is not a ``physical'' one in a sense that
the Nature is not obliged to be nice to us.}.

The number of {\em experimentally testable} scenarios is quite
limited. Basically, all the mechanisms for baryogenesis, which use new
physics at energies much higher than the Fermi scale do not fall in
this category. For example, the GUT baryogenesis (for a review see
\cite{Kolb:1983ni}) is not testable,  as it includes superheavy
particles with masses of the order of ${\cal O}(10^{16})$ GeV, the
direct search of which is impossible. The  thermal leptogenesis (for a
review see \cite{Davidson:2008bu}) is only partially testable. It did
pass the important tests coming from the observed neutrino masses and
mixings, but the prediction of the baryon asymmetry in it depends on
CP-breaking phases which cannot be measured in low energy experiments,
as the typical mass of the superheavy Majorana leptons is ${\cal
O}(10^{10})$ GeV - too high to allow their creation in laboratory. 

If new particles, responsible for BAU, are in the reach of existing or
future colliders, the baryogenesis mechanism has more chances to be
testable. One of the examples is the resonant leptogenesis of 
\cite{Pilaftsis:2004xx,Pilaftsis:2005rv}. It uses essentially the same
physics as thermal leptogenesis, but relies on a possible degeneracy
of heavy Majorana leptons. This allows to shift their masses to the
region accessible by the LHC or linear colliders, but still larger than
the electroweak scale. Though the interactions of one of the Majorana
leptons, relevant for BAU, are too weak to lead to its discovery at
colliders \cite{delAguila:2006dx}, two other heavy leptons can be
found and a peculiar flavour structure of their interactions can
provide the ``smoking gun'' signature of the resonant leptogenesis
\cite{Pilaftsis:2004xx,Bray:2007ru} \footnote{We thank A. Pilaftsis
for discussion of this point.}.

Yet another example is the electroweak baryogenesis (for reviews see 
\cite{Cohen:1993nk,Rubakov:1996vz,Trodden:1998ym}) where  the source
of baryon number non-conservation is the rapid high-temperature
anomalous processes with fermion number violation
\cite{Kuzmin:1985mm}, and departures from thermal equilibrium are due
to strongly first order electroweak phase transition 
\cite{Shaposhnikov:1986jp,Shaposhnikov:1987tw}. The electroweak
baryogenesis requires an extension of the Standard Model {\em right
above} the electroweak scale, leading to specific predictions for LHC
and ILC  (see \cite{Carena:2008vj,Chung:2008aya} for recent
discussions).

The baryogenesis can also occur due to new particles with the masses
{\em considerably smaller} than the Fermi scale. An example (actually,
the the only one, known to the authors) is associated  with the
minimal extension of the Standard Model (SM) in neutrino sector - the
$\nuMSM$ \cite{Asaka:2005an,Asaka:2005pn}. In this model one adds to 
the SM three singlet Majorana leptons with masses smaller than the
electroweak scale and uses the standard sea-saw Lagrangian, though
with the completely different choice parameters (for a review see 
\cite{Boyarsky:2009ix}). The lightest neutral singlet lepton with the
mass in ${\cal O}(10)$ keV region can be the dark matter (DM)
candidate; the other two with masses $M\sim {\cal O}(1)$ GeV would
then generate the baryon asymmetry of the Universe.  The Higgs boson
with non-minimal coupling to gravity  in the same model can play the
role of the inflaton and make the Universe flat, homogeneous and
isotropic, producing cosmological perturbations leading to structure
formation \cite{Bezrukov:2007ep}. The fact that so modest and natural
modification of the SM can address at once many problems which the SM
cannot solve (neutrino masses and oscillations, dark matter, baryon
asymmetry of the Universe and inflation) forces us to take it
seriously and study in detail the predictions of $\nuMSM$, which can
be used as a guideline for experimental searches for new particles it
contains\footnote{Further extensions of the $\nu$MSM are surely
possible. See 
\cite{Shaposhnikov:2006xi,Kusenko:2006rh,Anisimov:2008qs,Bezrukov:2009yw}
for discussion of the model with extra scalar singlet and
\cite{Bezrukov:2008ut} for analysis of the model with
higher-dimensional operators added. Of course, the predictions of
these models are less certain than those of the $\nu$MSM.}.  

The astrophysical observations coming from X-ray satellites tell that
the coupling of the lightest (DM) singlet fermion $N_1$ to active
neutrinos is so small that it practically does not contribute to the
see-saw formula \cite{Boyarsky:2006jm}. This means that two other
singlet fermions of the $\nuMSM$ $N_{2,3}$ must be able to explain
{\em simultaneously} the pattern of neutrino masses and oscillations
and lead to baryogenesis. The parameter counting goes as
follows\footnote{From now on we omit the DM singlet fermion from
consideration, as irrelevant for active neutrino masses and for
baryogenesis.}. Out of  11 new parameters of the model 7 can be
(potentially) measured in experiments with active neutrino flavours
only (these are 3 mixing angles, one Dirac CP-violating phase, one
Majorana phase, and 2 active neutrino masses, the lightest one is
massless in this approximation). The other 4 can be conveniently
chosen as $M=\frac{1}{2}(M_1+M_2)$ -- the average Majorana mass,
$\Delta M_M=\frac{1}{2}(M_1-M_2)$ -- the mass splitting, $\epsilon < 1$ -- the
ratio of the strengths of the couplings of $N_2$ versus  $N_3$ to
leptons, and finally $\eta$ -- an extra CP-violating Majorana phase.
Whether $N_{2,3}$ can be found experimentally depends crucially on 2
parameters -- their mass and $\epsilon$ \cite{Gorbunov:2007ak}. The
main aim of the present paper is to find the constraints on these two
parameters from the requirement of successful baryogenesis.  

The idea that a pair of light (with masses in GeV region) and almost
degenerate Majorana leptons can result in baryogenesis through their
oscillations goes back to ref. \cite{Akhmedov:1998qx}. The equations
governing baryogenesis in the $\nuMSM$,  incorporating the relevant
physical effects (singlet fermion oscillations, transfer of fermion
number from singlet species to active leptons and back, etc.) were
formulated in \cite{Asaka:2005pn}, where it was shown that the $\nuMSM
$ can lead to baryon asymmetry of the Universe and give a dark matter
candidate, being perfectly consistent with the data on neutrino
oscillations. In that paper a perturbative solution to the kinetic
equations was found, valid in a certain part of the parameter
space\footnote{It is required that both $N_{2,3}$ are out of
equilibrium at the sphaleron freezing temperature $T_{sph}$ and that
the rate of $N_{2,3}$ oscillations exceeds the rate of the Universe
expansion at $T_{sph}$.}. A semi-quantitative analytic consideration of
baryogenesis in the $\nuMSM$ beyond perturbation theory has been 
carried out in \cite{Shaposhnikov:2008pf}, where different essential
time scales were analysed, and the generic dependence of the baryon
asymmetry on some of the $\nuMSM$ parameters was elucidated.

An accurate prediction for properties of singlet fermions call,
however, for a complete exploration of the parameter space and thus
for a numerical solution of the kinetic equations, without any
assumptions about hierarchies of different time scales, made in
analytical computations. To identify the parameter-space allowing for
baryogenesis, we calculate the baryon asymmetry as a function of the
constants of the $\nuMSM$ as follows. We fix the parameters of the
neutrino mass matrix ($\theta_{12},~\theta_{23},~\theta_{13},~\Delta
m^2_{atm}$, and $\Delta m^2_{sol}$) to their experimental values. For
any given mass $M$, mass difference $\Delta M_M$ and $\epsilon$ we
determine the maximum of BAU as a function of unknown CP-violating 
phases and require that this value is larger than the observed one.
This allows to fix the 3-dimensional domain of parameters $(M,~\Delta
M_M,~\epsilon)$, most relevant for experimental searches of $N_{2,3}$.

This paper is organised  as follows. In Section \ref{se:basic} we 
summarise  the basic properties of the model and formulate the kinetic
equations  for baryogenesis. In Section \ref{se:baryon_asymmetry} we
describe the results of numerical analysis. The Section \ref{se:concl}
is conclusions. A number of appendixes contain details of the
numerical procedure. In fact, a direct numerical solution of the
kinetic equations poses a technical problem, as the corresponding
differential equations belongs to the so-called ``stiff'' type. The
reason is the existence of the many vastly different time scales,
associated with oscillations and with relaxations of distinct types of
deviations from thermal equilibrium. In appendix \ref{se:method} we
transform the kinetic equations, governing the evolution of the system
into a convenient form, which allows to elucidate the hierarchy of the
relevant time-scales (the rate of the singlet fermion oscillations,
coherence loss, transfer of leptonic number from sterile to active
species, back reaction, leptonic flavour non-conservation, etc) and
describe an effective way to solve them numerically. In appendix 
\ref{se:results:ver} we discuss the dependence of baryon asymmetry on
the parameters $M,~\Delta M_M$ and $\epsilon$, and its time evolution.
In \ref{se:max} we compute the maximal baryon asymmetry which can be
created in \nuMSM.

\section{Basic equations}
\label{se:basic}
This chapter is a brief review of the baryogenesis in the $\nuMSM$.
First,  we will describe the model, then we will formulate the kinetic
equations,  and finally we present the results of perturbative
computation of the  asymmetry, to be compared  with numerics later. 
We use the same  notations as in \cite{Shaposhnikov:2008pf}.

\subsection{Lagrangian and parameters}
\label{sse:lagr}
The part of the Lagrangian of the $\nu$MSM we are interested at can be 
written as:
\begin{eqnarray}
\label{Lagrangian-2008}
\mathcal{L}  & = & i \bar{N}_I \partial_\mu \gamma^\mu
N_I + \mathcal{F}
\Bigg(\frac{e^{-i\eta/2}}{\sqrt{\epsilon}} \bar{L}_2 N_2 +
\sqrt{\epsilon} e^{i\eta/2} \bar{L}_3 N_3 \Bigg) \tilde{ \Phi} 
\nonumber\\ & &  - M \bar{N}_2^c N_3  - \frac{\Delta M_M}{2}
(\bar{N}_2^c N_2 + \bar{N}_3^c N_3) + h.c., 
\end{eqnarray}
where $N_2$ and $N_3$ are the singlet leptons and $M \pm \Delta M_M $
their Majorana masses, $\tilde{\Phi}_i = \epsilon_{ij} \Phi^*_j$ is
the  Higgs doublet and $v = 174$ GeV its vacuum expectation value,
$\epsilon$ is the ratio between  the strengths of the couplings of $N_3$
and $N_2$ ($\epsilon<1$ by convention) and $\eta$ is a CP-violating
phase as  explained in the Introduction. The constant $\mathcal{F}^2$
is expressed through active neutrino masses $m_i$ as 
\be
\mathcal{F}^2  =  \frac{M}{2 v^2}\sum_i m_i.
\ee
The $6$ out of the $11$ new parameters of the model are explicitly 
written in the Lagrangian \eqref{Lagrangian-2008} ($M$, $\Delta M_M$, 
$\epsilon$, $\eta$ and $m_i$). The other $5$, related to the active 
neutrino mixing matrix, are  hidden is $L_2$ and $L_3$ which are 
combination of $L_e$, $L_{\mu}$  and $L_{\tau}$, the lepton doublets. 
These parameters are $3$ mixing  angles $\theta_{12}$, $\theta_{13}$, 
$\theta_{23}$, one Dirac phase $\phi$ and one Majorana phase. It will
be denoted by  $\alpha$  for the normal active neutrinos  mass
hierarchy and $\xi$ for the inverted one, see \cite{Strumia:2006db}
for convention. The relations  between $L_{2,3}$ and the flavour
eigenstates were found in  \cite{Shaposhnikov:2008pf}, we present them
here for $\theta_{13} = 0$ and  $\theta_{23} = \frac{\pi}{4}$, which
is within the experimental error bars  (for a review see
\cite{Strumia:2006db}).

For the normal hierarchy we have:
\begin{eqnarray}
\nonumber
L_2 & = & +a_1\frac{L_\mu-L\tau}{\sqrt{2}} + a_2 L_e + a_3\frac{L_\mu+L
\tau}{\sqrt{2}}~,\\
L_3 & = & -a_1 \frac{L_\mu-L\tau}{\sqrt{2}} - a_2 L_e + a_3\frac{L_\mu+L
\tau}{\sqrt{2}}~,
\label{L23norm}
\end{eqnarray}
where the coefficients are :
\begin{eqnarray}
\nonumber
a_1 & = & i e^{-i(\alpha+\phi)} \sin\varrho\cos\theta_{12}~,\\
\nonumber
a_2 & = & i e^{-i\alpha}        \sin\varrho\sin\theta_{12}~,\\
\label{normain}
a_3 & = & \cos\varrho~,\\
\nonumber
\tan\varrho & = & \sqrt{\frac{m_2}{m_3}}\simeq \left(\frac{\Delta m^2_{\rm 
sol}}{\Delta m^2_{\rm atm}}\right)^{\frac{1}{4}}\simeq 0.4~.
\end{eqnarray}

For the inverted hierarchy the expressions are similar :
\begin{eqnarray}
\nonumber
L_2 & = & +i e^{-i\phi} b_1 \frac{L_\mu-L\tau}{\sqrt{2}} + b_2 L_e ~,\\
L_3 & = & -i e^{-i\phi}b_2^* \frac{L_\mu-L\tau}{\sqrt{2}} + b_1^* L_e ~,
\label{L23inv}
\end{eqnarray}
with the coefficients :
\begin{eqnarray}
\nonumber
b_1 & = & \frac{1}{\sqrt{2}} \left[\cos\theta_{12}e^{-i\zeta}+i\sin\theta_{12}
e^{+i\zeta}\right]~,\\
\label{invmain}
b_2 & = & \frac{1}{\sqrt{2}} \left[\cos\theta_{12}e^{+i\zeta}+i\sin\theta_{12}
e^{-i\zeta}\right]~.
\end{eqnarray}\

For the analysis of kinetic equations, it is convenient to define the
Yukawa  coupling matrix $F_{\alpha I}$ between the singlet fermion and
the lepton  doublet in the following way:
\begin{eqnarray}
\sum_{\alpha} F_{\alpha 2} \bar{L}_{\alpha} & = & \frac{i \mathcal{F}}{\sqrt
{2}} \bigg( \frac{e^{-i\eta/2}}{\sqrt{\epsilon}} \bar{L}_2 -
\sqrt{\epsilon} e^{i
\eta/2} \bar{L}_3 \bigg), \nonumber\\
\sum_{\alpha} F_{\alpha 3} \bar{L}_{\alpha} & = & \frac{\mathcal{F}}{\sqrt
{2}} \bigg( \frac{e^{-i\eta/2}}{\sqrt{\epsilon }} \bar{L}_2 +
\sqrt{\epsilon} e^{i
\eta/2} \bar{L}_3 \bigg),
\end{eqnarray}\
which corresponds to a basis where the singlet fermion Majorana mass 
matrix $\mathcal{M}$ is diagonal and equals to:
\begin{eqnarray}
\mathcal{M} = \left( \begin{array}{ccc} M - \Delta M_M & 0 \\ 0 & M + \Delta 
M_M \end{array} \right).
\end{eqnarray}\

\subsection{The framework}
A convenient way to treat a system of interacting Majorana fermions
and neutrinos  is to use a density matrix $\rho$ \cite{Dolgov:1980cq,
Barbieri:1990vx, Sigl:1992fn}.  Since we consider $2$ singlet fermions
and $3$ active neutrinos and their anti-particles, $\rho$ is a $10
\times 10$ matrix. For baryogenesis the temperatures are above the
electroweak scale $T_W\sim 100$ GeV, we also take $M \ll T_W$. In this
case the matrix $\rho$ can be simplified \cite{Asaka:2005pn}: it can
be described by $3$  chemical potentials $\mu_{\alpha}$ for active
neutrino species and by  two $2 \times 2$ density matrices for the
sterile neutrinos, one for positive helicity states, $\rho_N$ and one
for negative helicity states $\bar{\rho}_N$. The evolution of the
system satisfies the kinetic equations which take into account
creation of singlet fermions, their oscillations, and generation of
lepton asymmetries \cite{Asaka:2005pn}. Introducing the CP-even and
CP-odd deviations from thermal equilibrium:
\begin{eqnarray}
\delta \rho_+ & = & \frac{\rho_N + \bar{\rho}_N}{2} - \rho^{eq}, 
\nonumber\\
\delta \rho_- & = & \rho_N - \bar{\rho}_N,
\end{eqnarray}
where $\rho^{eq}=exp \big(-\frac{p}{T} \big)$ with $p$ the momentum 
of the sterile neutrino  \cite{Asaka:2005pn}, these kinetic equations 
are \cite{Asaka:2005pn,Shaposhnikov:2008pf}:
\begin{eqnarray}
\label{Eq-of-motion-rhoS}
i \frac{d \delta \rho_+}{dt} & = & [\re H^{int}_N, \delta \rho_+] - \frac{i}{2} 
\{ \re \Gamma_N, \delta \rho_+ \} + i \frac{T}{8} \sin\varphi \Big(
F^{\dagger} \mu F -F^T \mu F^* \Big)\nonumber\\ & + & \frac{i}{2}
[\im H^{int}_N,\delta \rho_-]  + \frac{1}{4} \{ \im \Gamma_N, \delta \rho_- \}, \\
\label{Eq-of-motion-rhoA}
i \frac{d \delta \rho_-}{dt} & = & [\re H^{int}_N, \delta \rho_-] - \frac{i}{2} 
\{ \re \Gamma_N, \delta \rho_- \} + i \frac{T}{4} \sin\varphi \Big( F^{\dagger} 
\mu F +F^T \mu F^* \Big) \nonumber\\ & + & 2 i [\im H^{int}_N,\delta \rho_
+] + \{ \im \Gamma_N, \delta \rho_+ \}, \\
\label{Eq-of-motion-mu}
i \frac{d \mu}{dt} & = & \diag \bigg[- \frac{i}{2} \{ \Gamma_L,\mu \} 
\nonumber\\ & + & i \frac{T}{16} \sin\varphi \Big(F ( \delta \rho_- + 2 \delta 
\rho_+) F^{\dagger} - F^* ( \delta \rho_- - 2 \delta \rho_+) F^T \Big) \bigg].
\end{eqnarray}
Here  $F_ {\alpha I}$ is Yukawa coupling matrix introduced in the
previous chapter,
\be
\Gamma_N  =   \frac{T}{4} \sin\varphi \ F^{\dagger}F,~~~~
\Gamma_L  =  \frac{T}{4} \sin\varphi \ \re[F F^{\dagger}]
\label{rates}
\ee
are the rates of creation of singlet and active fermions
correspondingly, and the oscillation Hamiltonian is given by
\begin{eqnarray}
\label{hosc}
H^{int}_N = \frac{T}{8}  F^{\dagger}F + \frac{1}{T} \diag \big[-M \Delta
M_M, M \Delta M_M \big].
\end{eqnarray}
The expressions (\ref{rates}, \ref{hosc}) were found for temperatures
above $T_W$ in the symmetric  phase of electroweak theory, $\sin\varphi
\simeq 0.02$  is related to the ratio of the absorptive to the real
part of neutrino propagator in the medium
\cite{Akhmedov:1998qx,Asaka:2005pn}. 

The kinetic equations are written in the so-called monochromatic
approximation, which assumes that the different energy modes of
singlet fermions are effectively decouple. This is true in our case,
as the collisions between $N_{2,3}$ can be surely neglected, and the
problem is linear in terms of the density matrix for the singlet
fermions\footnote{In general, this is not the case for active
neutrinos in core collapse supernova, where one needs to take into
account neutrino-neutrino interactions, leading to non-linear effects
\cite{Dasgupta:2009mg,Duan:2010bg}. We thank E. Akhmedov and A.
Smirnov for discussion of this point.}.

The equations 
(\ref{Eq-of-motion-rhoS},\ref{Eq-of-motion-rhoA},\ref{Eq-of-motion-mu}), 
supplemented by initial conditions with no lepton asymmetry for the
active leptons ($\mu = 0$) and zero abundance for the singlet fermions
($\delta  \rho_+ = -\rho^{eq}$, $\delta \rho_- = 0$) describe the
evolution of active and sterile neutrinos from CP-symmetric initial
state. The choice of the initial condition for the CP-even part of the
density matrix  $\delta  \rho_+$ is motivated by the following facts.
First, the $\nu$MSM interactions of singlet fermions are so weak that
they are out of thermal equilibrium till temperature $T_+$ which is
close to the electroweak scale (see the estimates of $T_+$ as a
function of the parameters of the $\nu$MSM in 
\cite{Shaposhnikov:2008pf}). Therefore, they are not produced at
$T>T_+$. Second, in the minimal setup of the $\nu$MSM, where inflation
occurs due to the Higgs boson \cite{Bezrukov:2007ep}, the initial
conditions for the kinetic equations can be found. They coincide
exactly  with those formulated above \cite{Bezrukov:2008ut}. Moreover,
the dependence of the produced baryon asymmetry on the initial
condition for the CP-even part of the density matrix is linear in 
$\delta  \rho_+$.  Even if the singlet leptons had relatively strong
non-$\nu$MSM interactions at high energy scale, their initial (for our
kinetic equations) abundance will be in the interval  $-\rho^{eq} <
\delta  \rho_+ < 0$.  In other words, the produced baryon asymmetry
for generic initial conditions for $\delta  \rho_+$ can only be
smaller than what we found. So, the prediction of the domain of the
parameters, were baryogenesis in the $\nu$MSM {\em is not possible},
and therefore, identification of the experimental goals for their
search (Figs. \ref{Baryon-asymmetry-epsilon-deltam} - \ref{TauN}) is
robust. At the same time, if $|\delta  \rho_+| < \rho^{eq}$, the
region where  baryogenesis in the $\nu$MSM {\em is possible} will
decrease correspondingly.

An analytic solution to these equations in a particular limit,
discussed in the following subsection, was found in
\cite{Asaka:2005pn}. The qualitative features of this system of
kinetic equations have been explored in \cite{Shaposhnikov:2008pf}. We
will use  them to compute numerically the baryon  asymmetry of the
Universe as a function of the $\nuMSM$ parameters in Section
\ref{se:baryon_asymmetry}. Note that the transfer of the lepton
asymmetry to the baryon asymmetry because of sphalerons is not
included into eqns.
(\ref{Eq-of-motion-rhoS},\ref{Eq-of-motion-rhoA},\ref{Eq-of-motion-mu}),
this can be done analytically in a standard way (see, e.g.
\cite{Burnier:2005hp}).

\subsection{Analytical expression for baryon asymmetry}
\label{sse:lagr-ana-exp}
The equations \eqref{Eq-of-motion-rhoS}, \eqref{Eq-of-motion-rhoA}
and  \eqref{Eq-of-motion-mu} can be solved perturbatively for small
Yukawa  couplings, provided the following assumptions are satisfied
\cite{Asaka:2005pn}:
\begin{enumerate}[(i)]
\item \label{ana-exp-cond-1} {Singlet fermions are out of thermal 
equilibrium for all temperatures above the sphaleron freeze-out,  
$\Gamma_N \cdot t_W < 1$ with $\Gamma_N \simeq \sin\varphi \frac{T 
\mathcal{F}^2}{8 \epsilon},~t_W \simeq M_0/T_W^2$, where $M_0 = 7 
\cdot 10^{17}$ GeV.}
\item \label{ana-exp-cond-2} {The baryon asymmetry creation temperature 
$T_L \simeq (4 M \Delta M_M M_0)^{1/3}$ 
\cite{Akhmedov:1998qx,Asaka:2005pn} is much higher than the electroweak 
temperature, $T_L > T_W$.}
\item \label{ana-exp-cond-3} {The mass difference between two singlet 
fermions is much larger than the mass difference between active 
neutrinos, $\Delta M_M \gg  0.04$ eV for the normal hierarchy and $
\Delta M_M \gg 8 \cdot 10^{-4}$  eV for the inverted hierarchy.}
\end{enumerate}
The baryon to entropy ratio (the observed one is $\frac{n_B}{s} \simeq 
(8.4 - 8.9) \cdot 10^{-11}$) is given by
\begin{eqnarray}
\label{Baryonic-Asymmetry}
\frac{n_B}{s} \simeq 7 \cdot 10^{-4} \Tr(\delta \rho_-) |_{T_W}~,
\end{eqnarray}
where $Tr(\delta \rho_-) |_{T_W}$ can be found in \cite{Asaka:2005pn}
and can be conveniently parametrised as 
\begin{eqnarray}
\label{Asymmetry_approx}
Tr(\delta \rho_-) |_{T_W} & \simeq & 36 \cdot \delta_{CP} \cdot \Bigg
( \frac{\sin\varphi  \mathcal{F}^2 M_0}{8 \ \epsilon \ T_W} \Bigg)^3 \cdot 
\Bigg( \frac{T^3_W}{4 M \Delta M_M M_0} \Bigg)^{2/3}, \\
\label{deltaCP}
\delta_{CP} & = & \frac{2\epsilon^3}{\mathcal{F}^6} \sum_{I, \alpha} |F_
{\alpha I}|^2 
\im[F_{\alpha 3} [F^{\dagger}F]_{32}F^{\dagger}_{2 \alpha}]~.
\end{eqnarray}
Due to the conditions \eqref{ana-exp-cond-1} and 
\eqref{ana-exp-cond-2}, the first and second  brackets respectively
are always smaller than $1$. This formula was used recently to compute
the baryon asymmetry of the Universe in the \nuMSM, where CP-violation
is solely originated from the CP violation in the mixing matrix of active
neutrinos (i.e. for $\eta=0$) in \cite{Asaka:2010kk}.

Qualitatively, if $\Gamma_N \cdot t_W \gg 1$, the singlet fermions
equilibrate, and dilute the baryon asymmetry. If $T_L \ll T_W$, the
oscillations of singlet leptons had no time to develop and the
creation of baryon asymmetry is strongly suppressed
\cite{Shaposhnikov:2008pf}. In other words, the maximal asymmetry is
generated when  $\Gamma_N \cdot t_W \sim 1$ and $T_L \sim T_W$
simultaneously.

In general case, the conditions \eqref{ana-exp-cond-1}-
\eqref{ana-exp-cond-3} are not satisfied, and numerical solution of
the kinetic equations is necessary. The results of this analysis are
presented in the next Section.


\section{Domain of the parameters leading to baryon asymmetry}
\label{se:baryon_asymmetry}
In this chapter, we determine the 3-dimensional domain of the
parameters  $\epsilon$, $M$ and $\Delta M_M$ that can lead to the observed
baryon  asymmetry. It is found in the following way. The active
neutrino mass matrix parameters are fixed to be
\begin{eqnarray}
\label{Active_neutrinos_parameters_1}
\sin\theta_{12} & = & \sqrt{0.3}, \ \theta_{13} = 0, \ 
\theta_{23} = \frac{\pi}{4}, 
\nonumber\\
m_2 & = & \sqrt{\Delta m^2_{sol}} = 9 \ meV, \ m_3 = \sqrt{\Delta m^2_
{atm}} = 50 \ meV,
\end{eqnarray}
for the normal hierarchy and
\begin{eqnarray}
\label{Active_neutrinos_parameters_2}
m_2 = m_3 = \sqrt{\Delta m^2_{atm}},
\end{eqnarray}
for the inverted hierarchy. The variation of the parameters in 
(\ref{Active_neutrinos_parameters_1},\ref{Active_neutrinos_parameters_2})
within experimental error bars (shown, for instance, in 
\cite{Strumia:2006db}) does not lead to any significant changes in the
admitted parameter range. The unknown CP-violating phases are chosen
in such a way that baryon asymmetry is maximised (see appendix 
\ref{se:max}). They are close to the values given below, 
\begin{eqnarray}
\label{Phases}
\eta \simeq \phi \simeq  \frac{\pi}{2}, \ \alpha \simeq \frac{\pi}{2}, \ 
\xi \simeq \frac{\pi}{4},
\end{eqnarray}
which corresponds to the following Yukawa couplings ratio:
\begin{eqnarray}
\label{Yukawa-Couplings-Normal}
|F_{e 2}|^2 : |F_{\mu 2}|^2 : |F_{\tau 2}|^2 \simeq 1 : 11 : 11,
\end{eqnarray}
for the normal hierarchy and
\begin{eqnarray}
\label{Yukawa-Couplings-Inverted}
|F_{e 2}|^2 : |F_{\mu 2}|^2 : |F_{\tau 2}|^2 \simeq 48 : 1 : 1,
\end{eqnarray}
for the inverted hierarchy. The second set corresponds to the model I
of \cite{Gorbunov:2007ak} and the first set is somewhat between the
models II and III, described in \cite{Gorbunov:2007ak}. The sphaleron
freezing temperature is set at $T_W = 140$ GeV corresponding to the 
Higgs mass $m_H = 125$ GeV \cite{Burnier:2005hp}. We have also done
the computations for $T_W = 170$ GeV which correspond to $m_H = 200$
GeV, but the results are very close to the case $T_W = 140$ GeV.

With this set of parameters the kinetic equations 
\eqref{Eq-of-motion-rhoS}, \eqref{Eq-of-motion-rhoA} and 
\eqref{Eq-of-motion-mu} are solved numerically with the methods
described in appendix \ref{se:method}. The admissible domain of the
parameters is determined from the condition
$n_B/s|_{theoretical}>n_B/s|_{observed}$.  For masses above $\sim 10$
GeV the computation should not be trusted, as the kinetic equations
are only valid in the limit $M \ll T_W$. We expect that the admitted
region of the singlet fermion masses in the \nuMSM \ closes up at $M
\sim M_W$  \cite{Shaposhnikov:2008pf}. A number of intermediate
results, such as dependence of baryon asymmetry on $\epsilon,~M$ and
$\Delta M_M$, and its time dependence, can be found in appendix
\ref{se:results:ver}.

In the left panel of Fig. \ref{Baryon-asymmetry-epsilon-deltam} we
present the projection of the allowed region to the $\epsilon~-~M$ plane
for both neutrino mass hierarchies. Here, for every $\epsilon$ and $M$
the asymmetry is extremised with respect to $\Delta M_M$; in the interior
region the asymmetry is greater than the observed one, the boundary
corresponds to $n_B/s|_{theoretical}=n_B/s|_{observed}$. The baryon
asymmetry generation is possible for very light singlet fermions,  $M
\simeq 10$ MeV for the  normal hierarchy and $M \simeq 1$ MeV for
the inverted one. They correspond to small $\epsilon \sim {\rm few}
\times 10^{-4}$.
\FIGURE{
\centerline{
\includegraphics[width=0.45\textwidth]{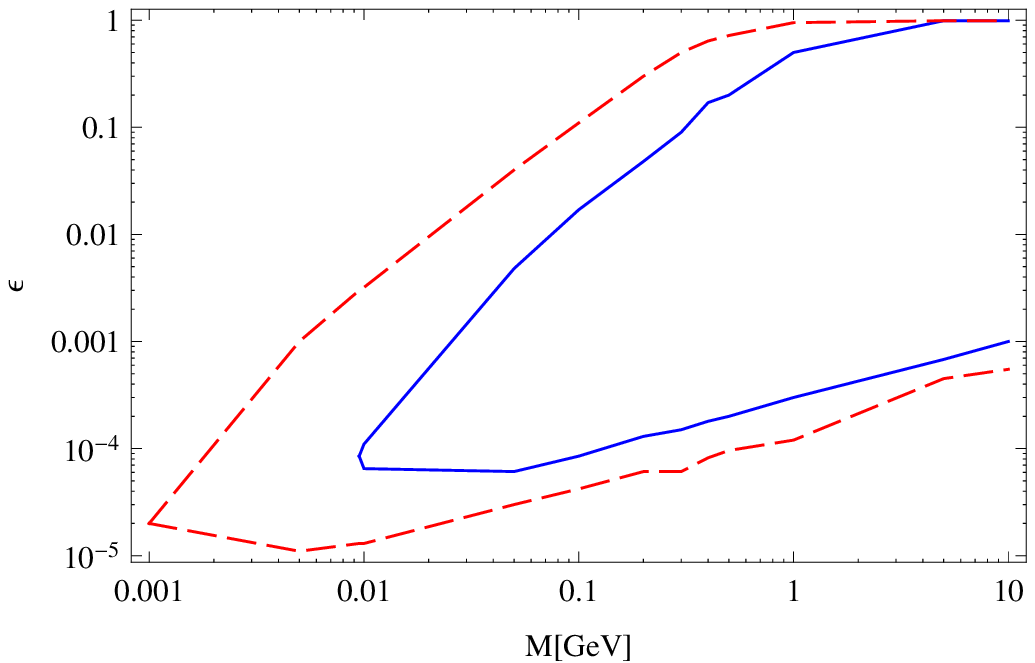} \ \ \ \
\includegraphics[width=0.45\textwidth]{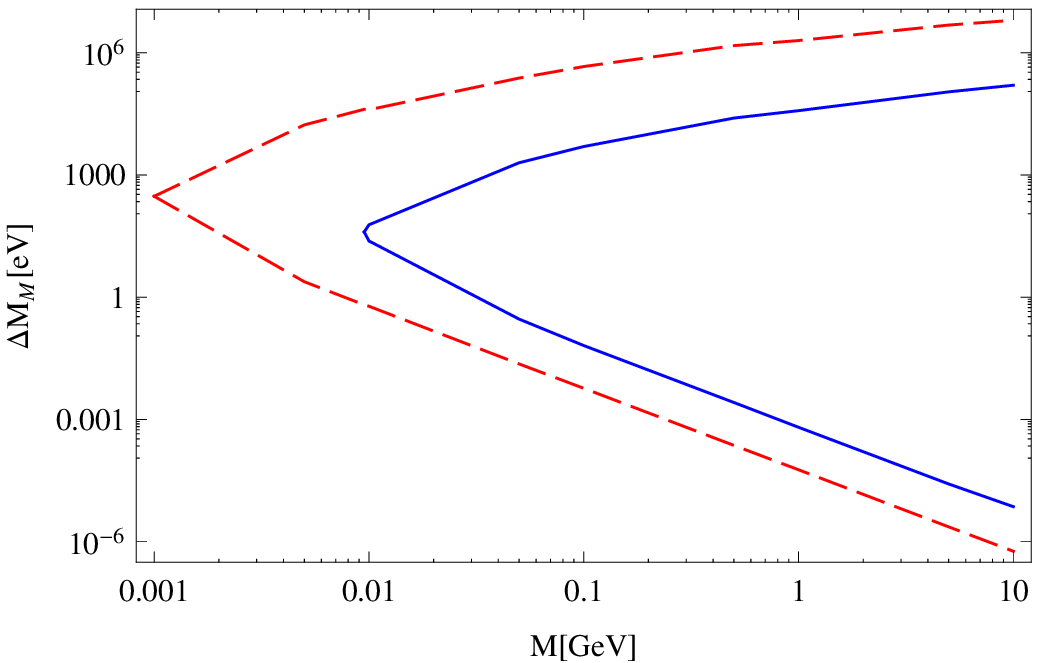}
}
\caption{Values of $\epsilon$ - $M$ (left panel) and $\Delta M_M$ - $M$ (right
panel) that leads to the observed baryon  asymmetry for the normal hierarchy
(blue - solid line) and for the inverted hierarchy  (red - dashed line).}
\label{Baryon-asymmetry-epsilon-deltam}
}

The right panel of Fig. \ref{Baryon-asymmetry-epsilon-deltam} gives
a projection of our domain to the $\Delta M_M~-~M$ plane, which is
derived in the similar way (asymmetry is now extremised with respect
to $\epsilon$ at fixed  $\Delta M_M$ and $M$). The admitted mass
difference ranges from a fraction of eV to MeV, depending on the
singlet fermion mass.

In Fig. \ref{Baryon-asymmetry} we present yet other projections to
the  $\Delta M_M~-~\epsilon$ plane for several singlet fermion masses
and two different types of the hierarchies. In general, for larger
masses the amount of the parameter space where baryogenesis happens is
larger. Also, the inverted hierarchy has more parameter space than the
normal one.  
\FIGURE{
\centerline{
\includegraphics[width=0.45\textwidth]{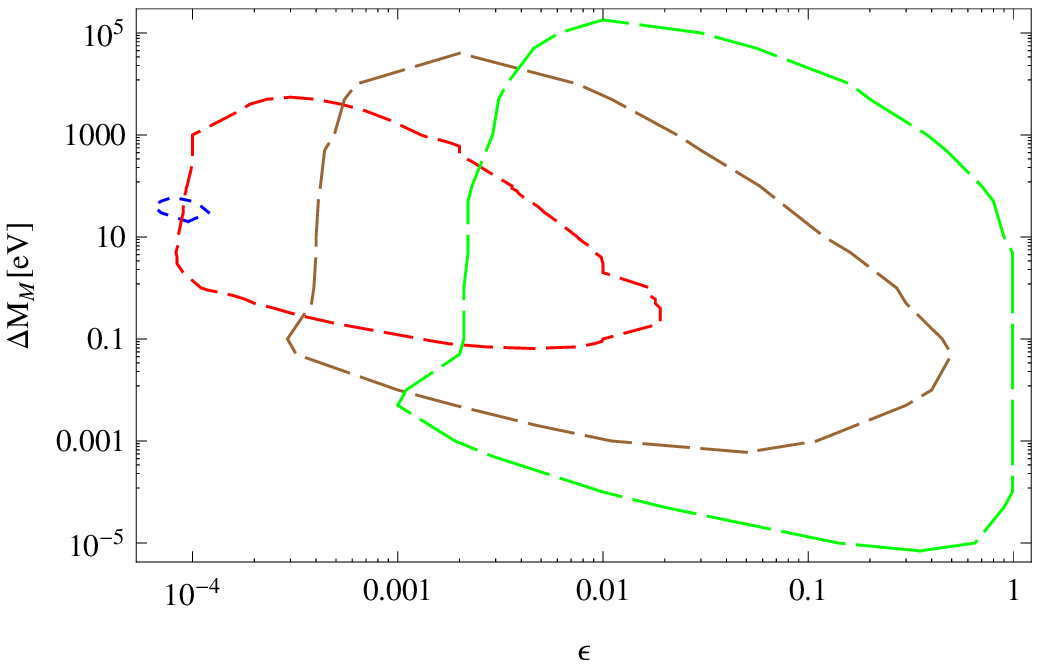}
 \ \ \ \
\includegraphics[width=0.45\textwidth]{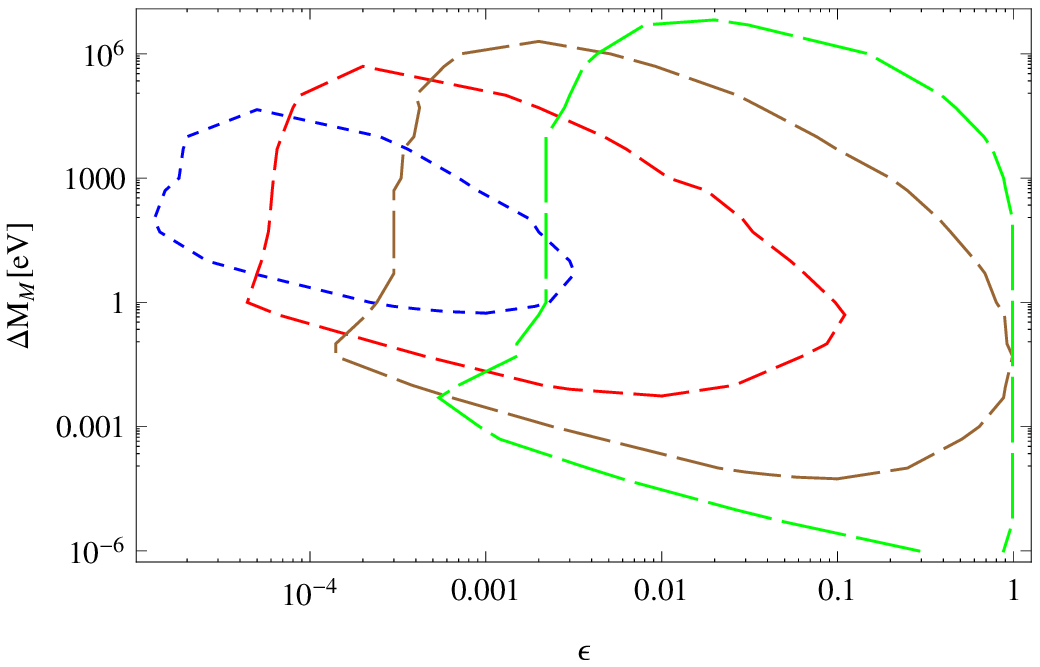}
}
\caption{Values of $\Delta M_M$ and $\epsilon$ that leads to the 
observed baryon asymmetry for different singlet fermion masses,  $M =
10,~100$ MeV, $1$, and $10$ GeV. The blue (shortest dashed) line
corresponds to $M = 10$ MeV, red (short dashed) - to  $M = 100$ MeV, 
brown (long dashed) -  $M = 1$ GeV and green (longest dashed) to $M =
10$ GeV. Left panel -  normal  hierarchy, right panel - inverted
hierarchy.}
\label{Baryon-asymmetry}
}

For most important parameters for experimental searches of the singlet
fermions are their mass and $\epsilon$, which determines the strength
of the coupling of $N_{2,3}$ to ordinary leptons. Usually it is
expressed through the mixing angle $U^2$ between ordinary and singlet
leptons (we take a sum over mixing angles with all active neutrino
flavours,  for an exact definition see \cite{Gorbunov:2007ak}). For
small $\epsilon$, 
\be
U^2 = \frac{\sum m_i}{4 M \epsilon}~.
\label{U2}
\ee
The region, where baryogenesis is possible in $U^2-~M$ plane is shown
in Fig. \ref{exp}, which is the main result of this 
paper\footnote{These figures supersede other similar  plots which
appeared previously in a number of  works \cite{Gorbunov:2007ak,
Boyarsky:2009ix} and in conference proceeding.}.  It is derived from
Fig. \ref{Baryon-asymmetry-epsilon-deltam} with the use  of
(\ref{U2}). We also plot there the exclusion regions coming  from
different experiments such as  BEBC \cite{CooperSarkar:1985nh},  CHARM
\cite{Bergsma:1985is}, and NuTeV \cite{Vaitaitis:1999wq}  and CERN
PS191 experiment \cite{Bernardi:1985ny,Bernardi:1987ek} (see also
discussion of different experiments in \cite{Atre:2009rg}).  For the
case of normal hierarchy, only  CERN PS191  have significantly entered
into cosmologically  interesting part of the parameter space of the
\nuMSM,  situated below the mass of the kaon. If the hierarchy  is
inverted, there are some constraints even for higher $N$ masses.  The
lower constraint on $U^2$, coming from baryon asymmetry of the
Universe, is somewhat stronger than the ``see-saw" constraint. In Fig.
\ref{TauN} we present the expected life-time of the singlet fermions
in an experimentally interesting region $M<2$ GeV.
\FIGURE{
\centerline{
\includegraphics[width=0.45\textwidth]{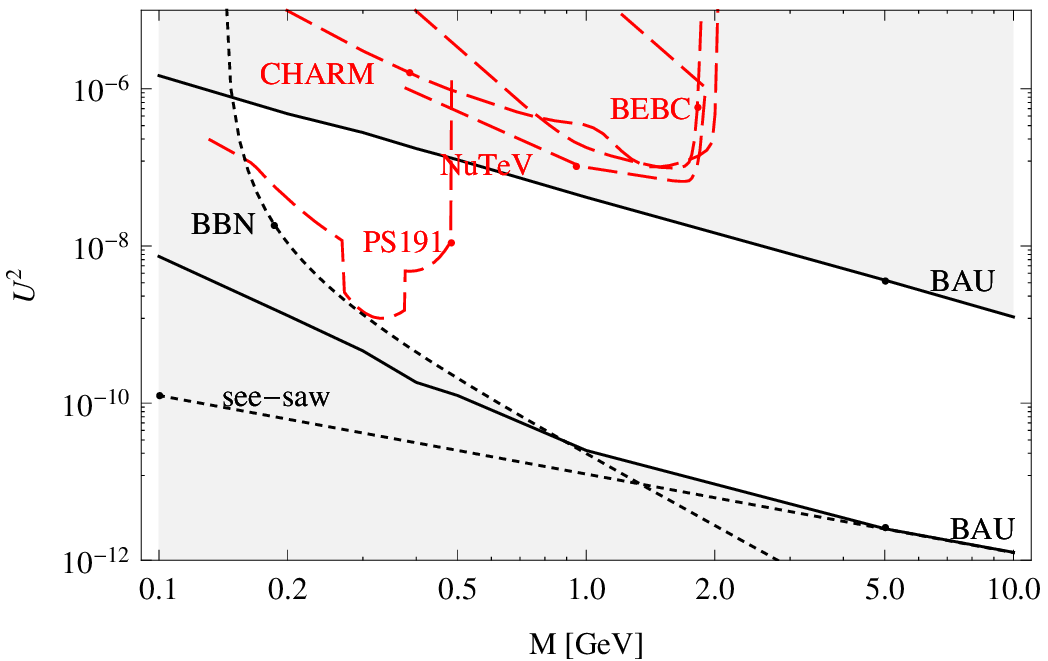}\ \ \ \
\includegraphics[width=0.45\textwidth]{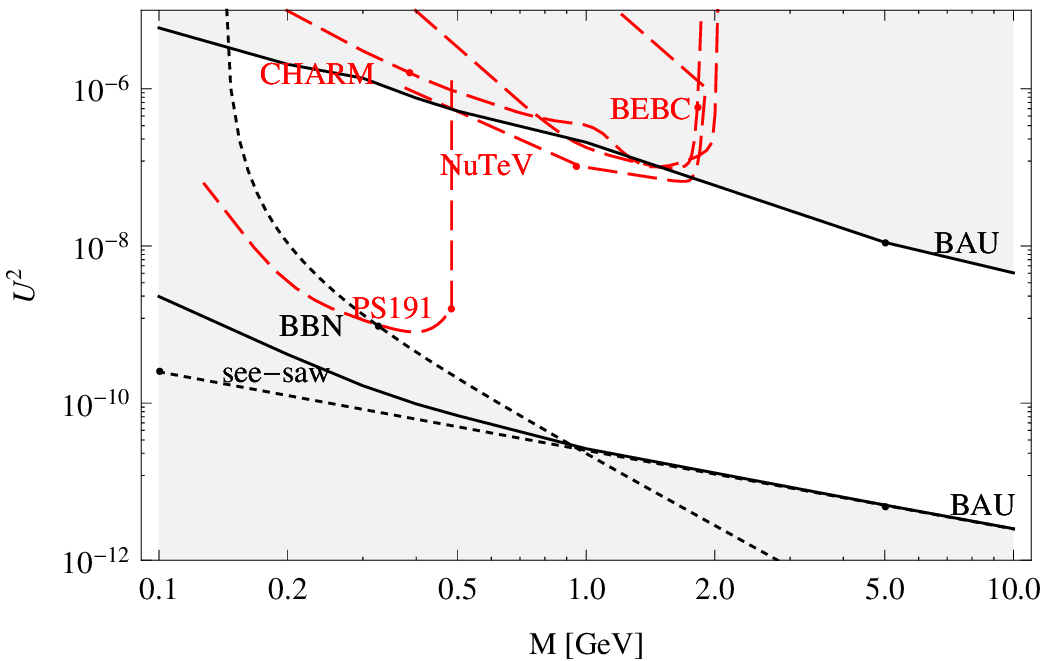}}
\caption{Constraints on $U^2$ coming from the baryon asymmetry of the
Universe (solid lines), from the see-saw formula (dotted line) and
from the big bang nucleosynthesis (dotted line). Experimental searched
regions are in red - dashed lines. Left panel -  normal  hierarchy,
right panel - inverted hierarchy.}
\label{exp}
}
\FIGURE{
\centerline{
\includegraphics[width=0.45\textwidth]{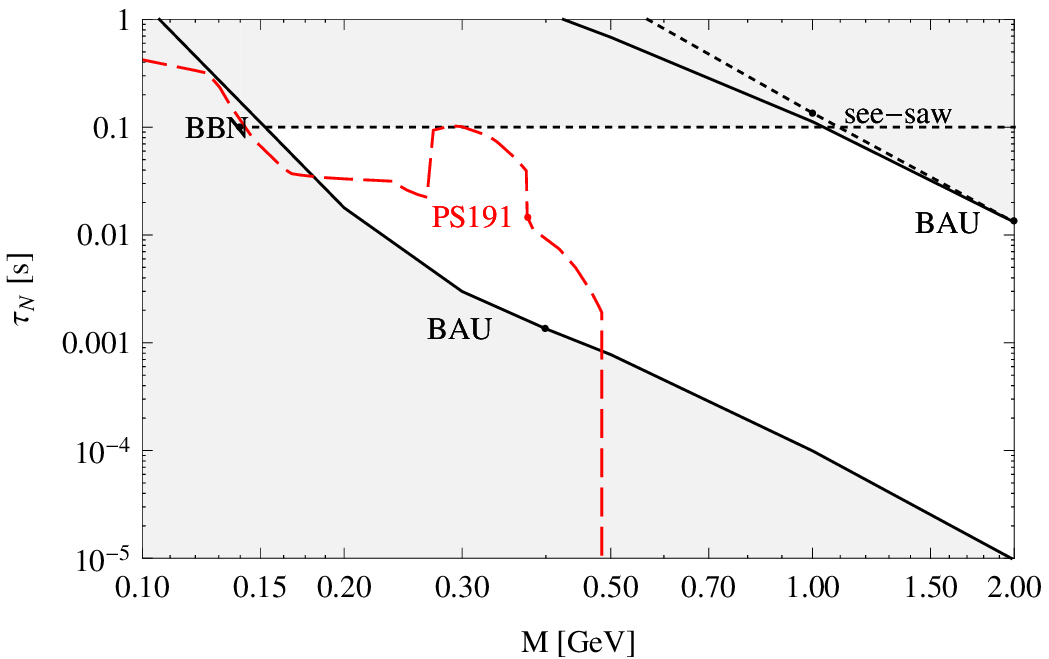}\ \ \ \
\includegraphics[width=0.45\textwidth]{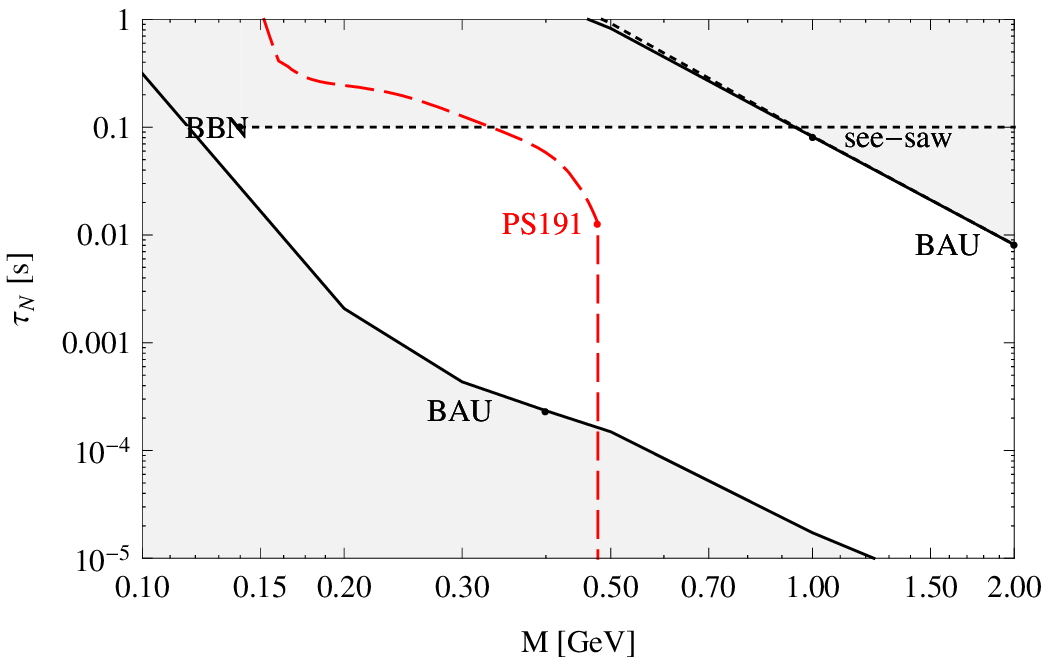}}
\caption{Constraints on the lifetime $\tau_N$ coming from the baryon
asymmetry of the Universe (solid lines), from the see-saw formula
(dotted line) and from the big bang nucleosynthesis (dotted line).
Experimental constraints from PS 191 are shown in red - dashed lines.
Left panel - normal  hierarchy, right panel - inverted hierarchy.}
\label{TauN}
}

A detailed discussion of possible experiments and signatures of 
neutral leptons leading to BAU generation in the $\nu$MSM can be found
in  \cite{Gorbunov:2007ak}. For the reader convenience, we summarise
below the main conclusions of this work.

Several distinct strategies can be used for the experimental search of
these particles. The first one is related to $N$ production ($U^2$
effect). The singlet fermions  participate in all the reactions the
ordinary neutrinos do with a probability suppressed roughly by a
factor $U^2$. Since they are massive, the kinematics of, say, two body
decays $K^\pm \rightarrow \mu^\pm N$, $K^\pm \rightarrow e^\pm N$ or
three-body decays  $K_{L,S}\rightarrow \pi^\pm + e^\mp + N_{2,3}$
changes  when $N_{2,3}$ is replaced by an ordinary neutrino.
Therefore, the study of  {\em kinematics} of rare $K,~D$, and $B$
meson decays can constrain the strength of the coupling of heavy
leptons.  The precise study of kinematics of rare meson decays is
possible in $\Phi$ (like KLOE), charm, and B factories, or in
experiments with kaons where their initial 4-momentum is well known
(like NA48, NA62 or E787 experiments).

The second strategy is to use the proton beam dump ($U^4$ effect). As
a first step, the proton beam heating the fixed target creates $K,~ D$
or $B$ mesons, which decay and produce $N_{2,3}$. The second step is a
search for decays of $N$ in a near detector, looking for the processes
``nothing" $\rightarrow$ leptons and hadrons.   To this end, quite a
number of already existing or planned neutrino facilities (related,
e.g., to CERN SPS, MiniBooNE, MINOS, J-PARC or LBNE at Fermilab),
complemented by a near {\em dedicated} detector can be utilised.

For the mass interval $M < M_K$, both strategies can be used. If
$m_K < M < m_D$, the search for the missing energy signal,
potentially possible at beauty, charm, and $\tau$ factories, is
unlikely to gain the necessary statistics and is  impossible at
hadronic machines like the LHC. Thus, the search for decays of neutral
fermions is the most effective opportunity.  The  dedicated
experiments on the basis of the proton beam NuMI or NuTeV at FNAL, SPS
at CERN, or J-PARC  can touch a very interesting parameter range for
$M < 1.8$ GeV. Experiments like NuSOnG \cite{Adams:2008cm} and 
HiResM$\nu$ \cite{Mishra:2008nx} should be able to enter in a
cosmologically interesting region for masses and mixing angles of
singlet fermions.

Going above $D$-meson but still below $B$-meson thresholds is very
hard if not impossible with the present or planned proton machines or
B-factories. To enter into a cosmologically interesting parameter
space would require the increase in the present intensity of, say,
CERN SPS beam by two orders of magnitude or to produce and study the
kinematics of more than $10^{10}$ B-mesons.  For $M$ above the
beauty threshold $\simeq 5$ GeV the baryogenesis in the $\nu$MSM
becomes untestable.

\section{Conclusions}
\label{se:concl}
In this paper we found the 3-dimensional domain of \nuMSM \ parameters
$(M,~ \Delta M_M,~\epsilon)$ that may lead to observed baryon
asymmetry of the Universe. It ranges from the singlet fermion
masses in MeV region to tens of GeV and from mass differences from a
fraction of eV to MeV. Thus, baryogenesis is a generic consequence of
the \nuMSM \ with nearly degenerate Majorana fermions. Our results can
be used to determine an ultimate goal for searches for cosmologically
interesting heavy neutral leptons, responsible for neutrino masses and
oscillations, and for baryon asymmetry of the Universe.

The analysis of our work can be improved in several aspects. In
particular, the kinetic equations we used are only approximate. They
are dealing with a typical particle in the plasma with momentum $\sim
T$. Clearly, the exact kinetic equations (to be derived yet) must have
integro-differential character. We also did not take into account the
mass corrections ${\cal O}\left(\frac{M_W^2}{T^2}\right)$ and ${\cal
O}\left(\frac{M}{T^2}\right)$. We postpone, however, the study of
these effects till the moment when the singlet fermions are
discovered.

\section*{Acknowledgements} 
This work was supported in part by the Swiss National Science
Foundation. We thank Takehiko Asaka and Dmitry Gorbunov for reading of
the manuscript and for useful comments.


\appendix
\section{The numerical method}
\label{se:method}
The straightforward numerical solution of equations
\eqref{Eq-of-motion-rhoS},  \eqref{Eq-of-motion-rhoA} and
\eqref{Eq-of-motion-mu} causes no problems for quite a wide range of
the parameters of the \nuMSM. However, in some corners of the
parameter space, where the damping is strong and/or oscillations are
very rapid, it is faced with a number of difficulties. Due to
simultaneous presence of many different time scales, related to
damping of various quantum numbers and due to oscillations between
singlet fermions, the  corresponding differential equations belongs to
the so-called ``stiff'' type. For example, for typical parameters $T_W
\simeq 100$ GeV, $M = 1$ GeV $\Delta M_M = 10^{-6}$ GeV and $ \epsilon =
0.1$,  the oscillation rate $\Delta E \simeq \frac{2 M \Delta  M_M}{T}
= 2 \cdot 10^{-8}$ GeV is some 7 orders of magnitude larger than the
rate of the singlet fermion creation, $ \Gamma_N \simeq \sin\varphi
\frac{T}{8 \ \epsilon} \mathcal{F}^2 = 2 \cdot  10^{-15}$ GeV.  So, to
solve kinetic equations numerically we will use a method which is very
similar to the one developed in  the Appendix A of Ref.
\cite{Farrar:1993hn}. The basic idea is to determine the damping and
oscillation rates at any given moment of time and then disentangle the
kinetic evolution accordingly. This is explained in detail below.

The system of equations \eqref{Eq-of-motion-rhoS},
\eqref{Eq-of-motion-rhoA} and \eqref{Eq-of-motion-mu} can be rewritten
in the form
\be
\label{Eq-for-new-method}
- \frac{1}{i} \frac{d \psi}{dt} = \mathcal{H} \psi~,
\ee
where $\psi$ is a real vector of dimensionality $11$, which includes
all unknowns, namely $2\times4$ real functions determining  density
matrices for singlet fermions, $\delta\rho_+$ and $\delta\rho_-$, and
$3$ chemical potentials for active leptonic asymmetries $\mu_\alpha$.
The $11\times11$ complex (but not Hermitian) matrix $\mathcal{H}$ can be
called effective Hamiltonian. The (complex, in general) eigenvalues of
$\mathcal{H}$ at some moment of time $t$ give the instant relaxation
and oscillation rates in the system.

Let us look for solutions of equation \eqref{Eq-for-new-method} in the
form:
\begin{eqnarray}
\label{Psi-new-method}
\psi (t) =e(t) E(t) \psi_0 (t),
\end{eqnarray}
where $\psi_0 (t)$ is some vector, and $e(t)$ is a matrix build
from the eigenvectors of the matrix $\mathcal{H}$:
\begin{eqnarray}
\label{eigenvectors}
\mathcal{H}e=ep,
\end{eqnarray}
where $p$ is a diagonal matrix,  constructed from eigenvalues of
$\mathcal{H}$  and the matrix $E(t)$ is defined by:
\be
\label{Exp-eigenvectors}
E(t) = \exp \Big(-i \int_{0}^{t} dt' p(t') \Big).
\ee
In this way the rapidly oscillating or exponentially damped behaviour
of $\psi (t)$ is factored out, and $\psi_0 (t)$ is slowly varying. Now
we will derive a convenient form of equation for $\psi_0 (t)$.

Let us denote the eigenvectors of $\mathcal{H}$ by $e_i$. Then we have
$\mathcal{H}e_i= p_j e_i$ and $e=(e_1,e_2,...)$. It is useful to
introduce another set of the eigenvectors $f_j$ defined  by $f_j
\mathcal{H}= f_j q_j$ and introduce the matrix $f$ as $f= 
\Bigg(\begin{array}{ccc} f_1 \\ f_2 \\ ... \end{array} \Bigg)$. The
sets of eigenvalues $q_j$ coincide with the set $p_i$ up to
permutation, since they are found from one and the same equation,
$det(\mathcal{H}-p)=0$ or $det(\mathcal{H}-q)=0$. If $p_i\neq q_j$, the
vectors $e_i$ and $f_j$ are orthogonal,  $f_j e_i = 0$. Therefore, the
reshuffling of the set $f_i$ can ensure that the matrices $f$ and $e$
satisfy $f e = A$, where $A$ is a diagonal matrix.  We can choose the 
normalisation of $f$ and $e$ in such a way that $A = Identity$, but we
will keep it arbitrary in what follows.

Using the orthogonality of $e$ and $f$ one can get an equation for
$\psi_0 (t)$:
\begin{eqnarray}
\label{equation-psi0-a}
\frac{d \psi_0}{dt} = - (AE)^{-1} f \frac{de}{dt} E \psi_0,
\end{eqnarray}
and the time evolution for the eigenvectors $e(t)$ and the eigenvalues
$p(t),~f(t)$: 
\begin{eqnarray}
\label{equation-p-a}
\frac{dp_i}{dt} & = & \frac{1}{A_{ii}} \Big(f \frac{d\mathcal{H}}{dt} e \Big)_
{ii}, \\
\label{equation-e-a}
\frac{de}{dt} & = & e \ G, \\
\label{equation-f-a}
\frac{df}{dt} & = & -  A\ G \ A^{-1} \ f, \\
\label{equation-G-a}
G_{ii} & = & 0, \ G_{ij} = \frac{1}{(p_j-p_i) A_{ii}} \Big(f \frac{d\mathcal{H}}
{dt} e \Big)_{ij}.
\end{eqnarray}

The equation (\ref{equation-psi0-a}) is still not suitable for
numerics due to explicit presence of exponentially small terms in the
matrix $E(t)$, coming from imaginary parts of the eigenvalues and
corresponding to strong damping (entering to thermal equilibrium of
the corresponding processes). To avoid this kind of  problem, we
remove the imaginary part of the eigenvalues from the exponential by
the change of variables:
\begin{eqnarray}
\label{change-of-variable}
\psi_0 (t) & = & \exp\Big(- \int_{0}^{t} dt' \im(p(t')) \Big) \tilde{\psi} (t), \\
\tilde{E} (t) & = & \exp\Big(-i \int_{0}^{t} dt' \re(p(t')) \Big),\\
\psi (t) & = &  e(t) \tilde{E}(t) \tilde{\psi}(t)~.
\end{eqnarray}
The new equation for $\tilde{\psi} (t)$ is:
\begin{eqnarray}
\label{equation-psi0-tilde}
\frac{d \tilde{\psi}}{dt} = \im(p) \tilde{\psi} - 
(A\tilde{E})^{-1} f \frac{de}{dt} 
\tilde{E} \tilde{\psi}.
\end{eqnarray}

These equations were used for compute the behaviour of the density
matrices and of the chemical potentials with the Wolfram Mathematica
command NDSolve. It is convenient to use a dimensionless variable $z$,
related to temperature as

\begin{eqnarray}
\label{z-tau}
z = \frac{h_0 \sin\varphi M_0}{4 T} =
\frac{h_0 \sin \varphi M_0}{4 \sqrt{M_0/ 2 t}},
\end{eqnarray}
with $h_0 = 2 \times 10^{-14}$ and changing from $0$ (initial state)  to $\sim 1$
(sphaleron freezing).

For some choice of the parameters the computation crashes and further
refinements of the method were necessary. 

First, if two of eigenvalues are equal to each other at some moment
$\tau_0$, the equation \eqref{equation-G-a} gets singular. To go
around this problem, we  continued the coordinate $z$ to the complex
plane, to go around the  singularity: $z = \tau + i
\frac{a}{\cosh^2(b(\tau-\tau_0))}$, where $\tau$,  $a$ and $b$ are some
real parameters. The result must be independent of the specific choice
of  $a$ and $b$, which we verified numerically.

Second, if the frequency of oscillations between two sterile neutrinos
is very high, the numerical solution of (\ref{equation-psi0-tilde}) is
extremely slow, as a very small time step size is required. At the
same time, no asymmetry is produced in this regime, exactly due to the
fact that the oscillations are rapid
\cite{Akhmedov:1998qx,Asaka:2005pn,Shaposhnikov:2008pf}. Therefore, we
introduced an artificial damping of oscillations, effectively
averaging them to zero. This was done by multiplying $\tilde{E}$ by 
$\exp \Big(-A \Big(\int_{0}^{t} dt' \re(p(t'))\Big)^2 \Big)$, where $A$
is some real parameter. We verified numerically that this procedure
indeed works.
%
%
\section{Baryon asymmetry as a function of \nuMSM \ parameters}
\label{se:results:ver}
In this appendix we present some numerical results which clarify the
dependence of the baryon asymmetry on the parameters of the model, and
consider its time evolution. A comparison with a perturbative
computation is also made. 

For computations, we set the parameters related to the active
neutrinos mass matrix as specified in eqns.
(\ref{Active_neutrinos_parameters_1},\ref{Active_neutrinos_parameters_2},
\ref{Phases}) and $T_W = 100$ GeV.

In Fig. \ref{BaryonAsymmetry-M} we present the dependence of asymmetry
on the mass $M$  for some choices of other parameters  ($\epsilon$ and
$\Delta M_M$).
\FIGURE{
\centerline{
\includegraphics[width=0.45\textwidth]{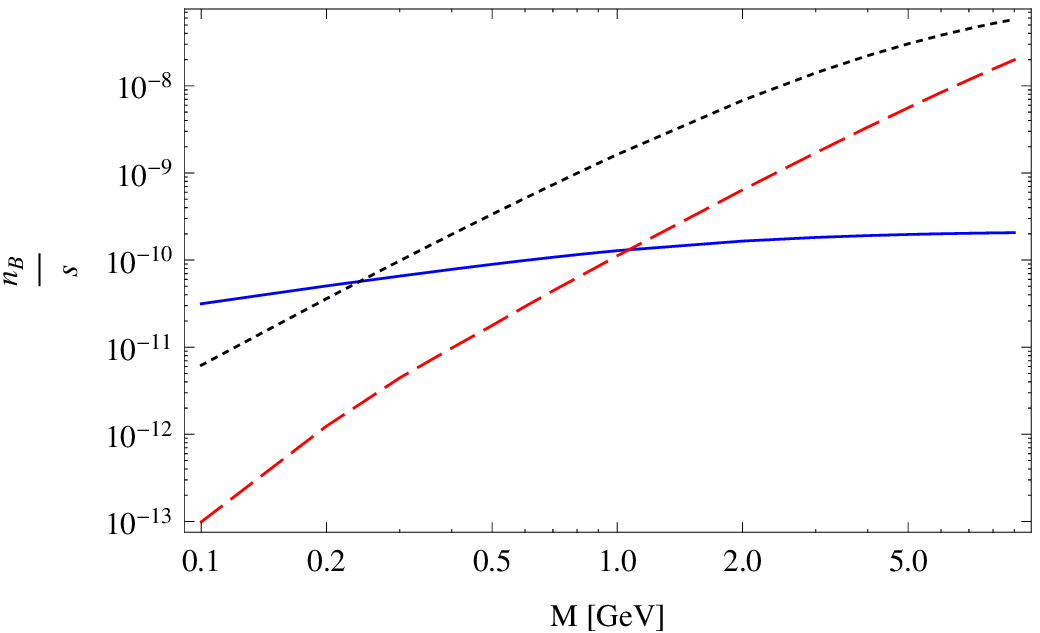}
}
\caption{Baryon asymmetry as a function of $M$ for $\epsilon = 5 \cdot
10^{-4}$, $\Delta M_M = 10^{-10}$ GeV in blue - solid line, for $\epsilon
= 0.5$, $\Delta M_M = 10^{-10}$ GeV in red - dashed line and for
$\epsilon = 7 \cdot 10^{-2}$, $\Delta M_M = 10^{-9}$ GeV in black -
dotted line}
\label{BaryonAsymmetry-M}
}

In Fig.~\ref{Asymmetry-DeltaM-Epsilon-M}a, we show the dependence of
the asymmetry on $\Delta M_M$, keeping $M = 2$ GeV and $\epsilon = 0.5$,
with latter two numbers chosen ad hoc. The upper line corresponds to
the perturbative solution given by \eqref{Asymmetry_approx}, and the
lower one to numerical integration. These  two curves are almost the
same for $\Delta M_M > 10^{-11}$ GeV. For smaller $\Delta M_M$ the
conditions (ii) and (iii) formulated in Section \ref{sse:lagr-ana-exp}
are not valid any more. The asymmetry decreases for very small $\Delta
M_M$, as the resonance has no time to develop.

In Fig.~\ref{Asymmetry-DeltaM-Epsilon-M}b we fix $M = 2$ GeV and  
$\Delta M_M = 10^{-6}$ GeV and vary $\epsilon$. The analytical
perturbative solution is close to the numerical one for $\epsilon >
0.01$. This is what is expected: the condition (i) of Sect.
\ref{sse:lagr-ana-exp} requires  $\epsilon \gsim  0.015$ to insure
that the sterile neutrinos are out of thermal equilibrium. For smaller
$\epsilon$ the asymmetry decreases because the singlet fermions start
to equilibrate.

In Fig.~\ref{Asymmetry-DeltaM-Epsilon-M}c we fix $\epsilon = 0.5$,
$\Delta M_M = 10^{-6}$ GeV and vary $M$. The curves start to deviate
from each other at large masses, where the condition (i) is not valid any
longer.
\FIGURE{
\centering
\begin{tabular}{cccc}
\includegraphics[width=0.45\textwidth]{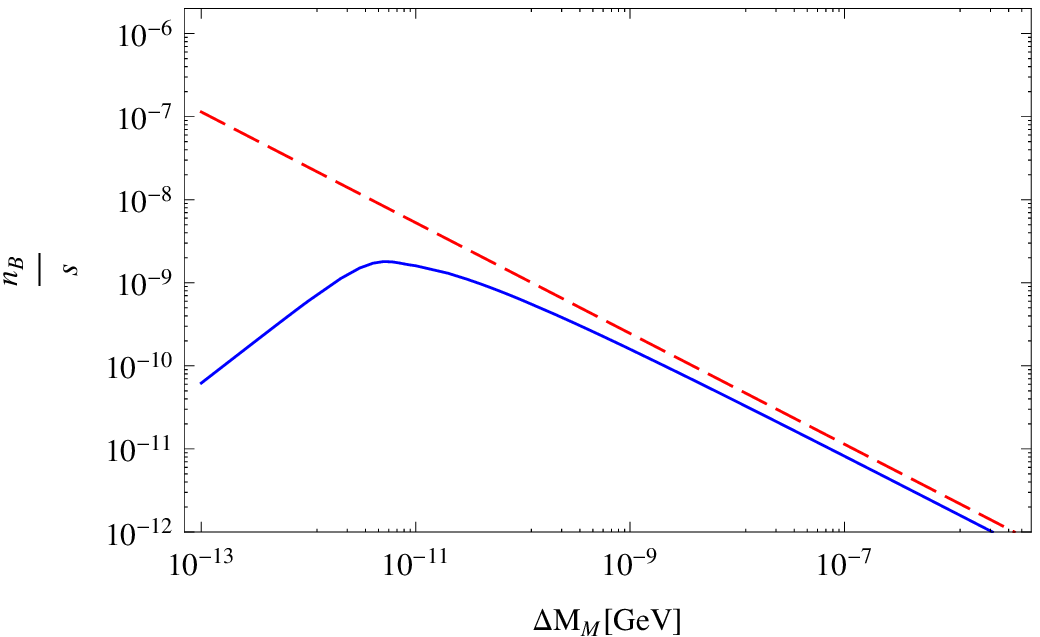} 
&
\includegraphics[width=0.45\textwidth]{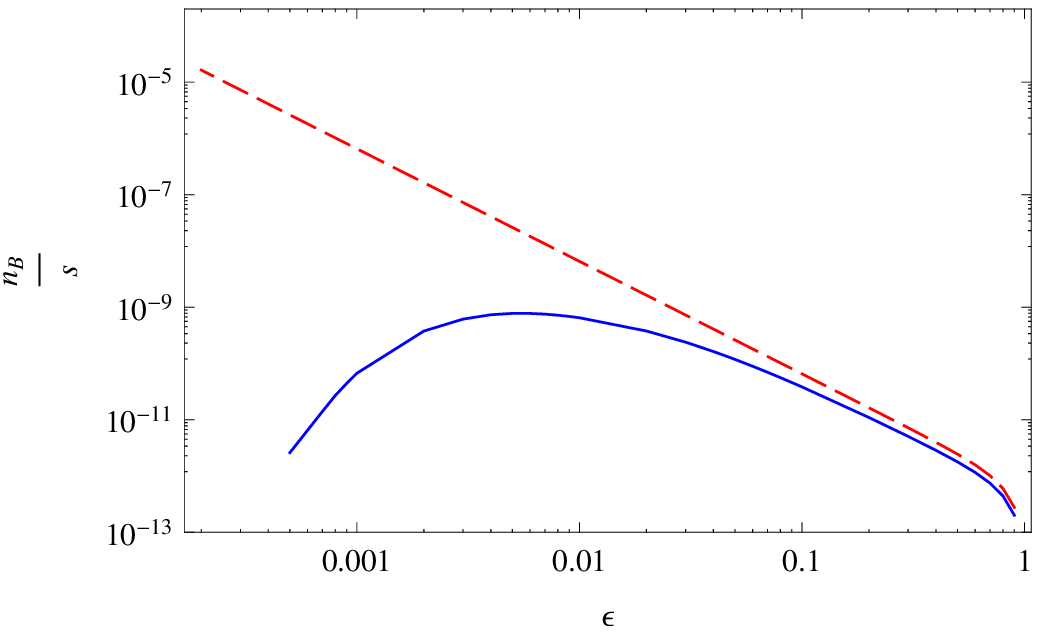} \\
(a) & (b) \\
\includegraphics[width=0.45\textwidth]{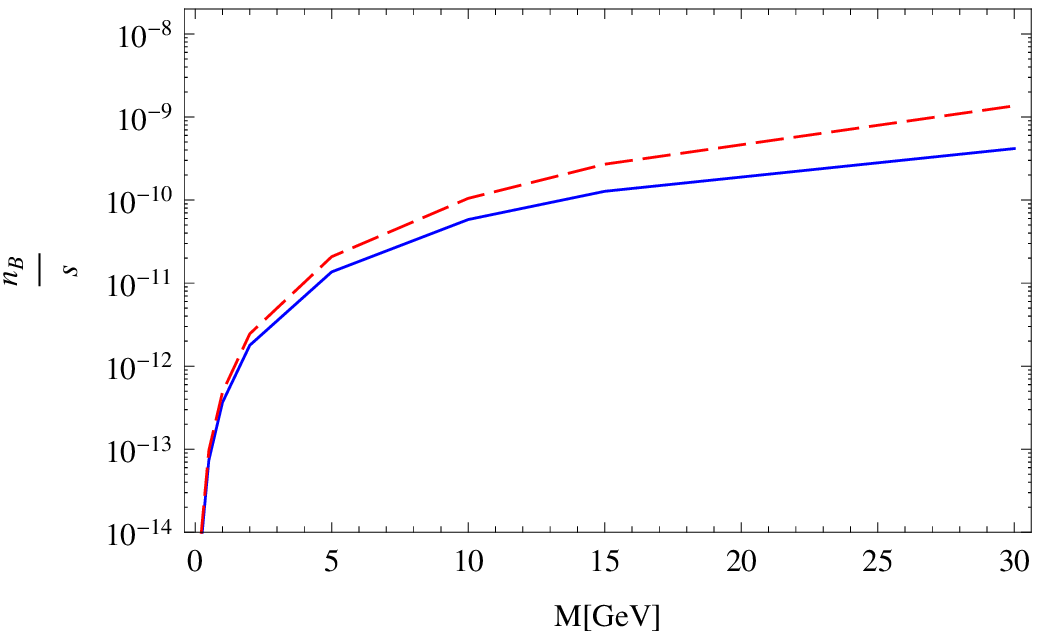} \\
(c)
\end{tabular}
\caption{Lepton asymmetry in function of $\Delta M_M$ (a), $\epsilon$ (b), 
$M$ (c), blue / solid line for numerical solution and red / dashed line for 
the analytic expression.}
\label{Asymmetry-DeltaM-Epsilon-M}
}

In Fig. \ref{Asymmetries} we show the time evolution of different
asymmetries for the following choice of parameters (for the  normal
hierarchy of neutrino masses). We take $\epsilon = 0.015$, $\Delta M_M =
10^{-8}$ GeV and $M = 1$ GeV, and consider evolution of the system
till temperature $T = 100$ GeV. These parameters are chosen in such a
way that equilibration of singlet fermions occurs at $T>T_W$ (so that
$\Gamma_N t_W >1$). Therefore, one would expect that the asymmetry
start to decrease after $t \simeq 1/\Gamma_N$. Also, for this choice
$T_L > T_W$ (see Sect. \ref{sse:lagr-ana-exp}), indicating that the
production of asymmetry in $N_{2,3}$ occurs around the time
corresponding to $T_L$. The time variable $\tau$ used in these  plots
is defined in appendix \ref{se:method}.
\FIGURE{
\centering
\begin{tabular}{cccc}
\includegraphics[width=0.45\textwidth]{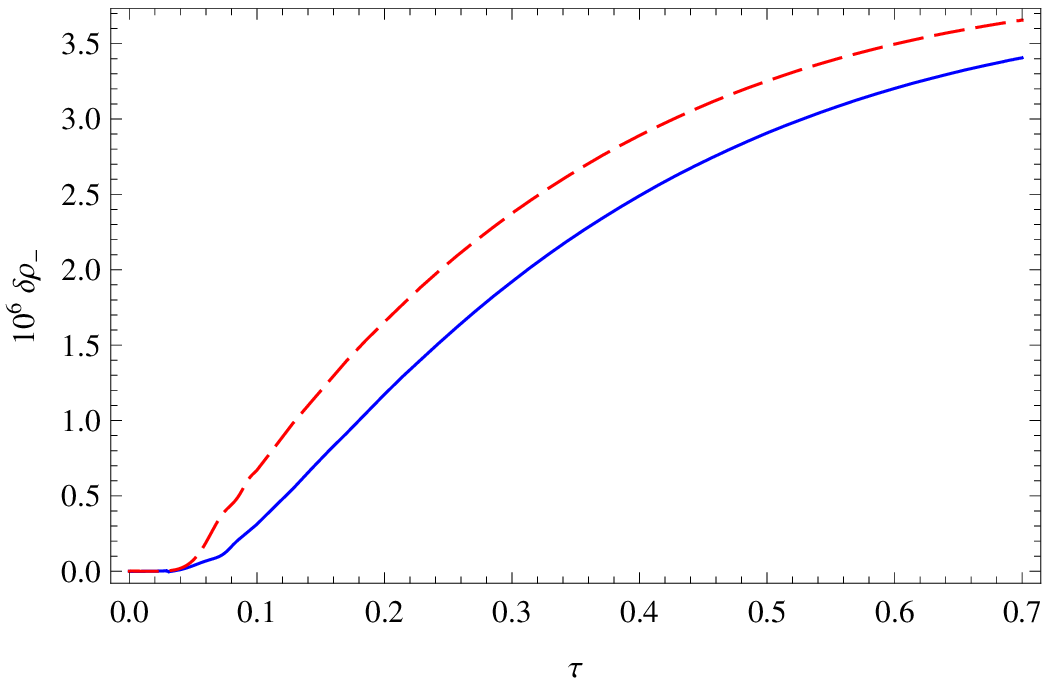} 
&
\includegraphics[width=0.45\textwidth]{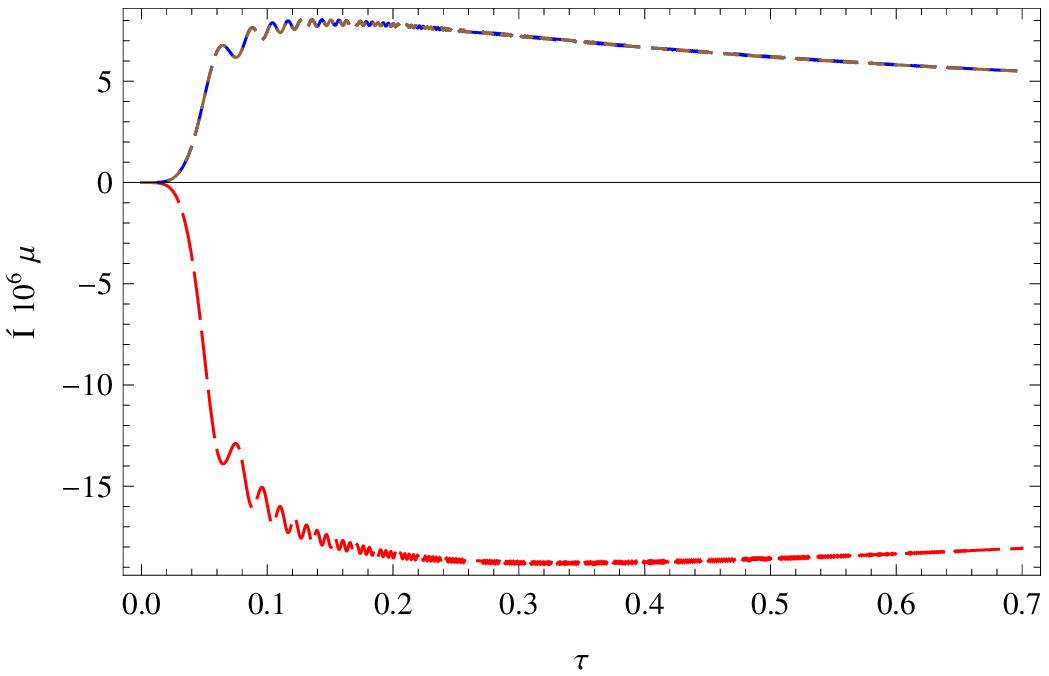} 
\\
(a) & (b) \\
\includegraphics[width=0.45\textwidth]{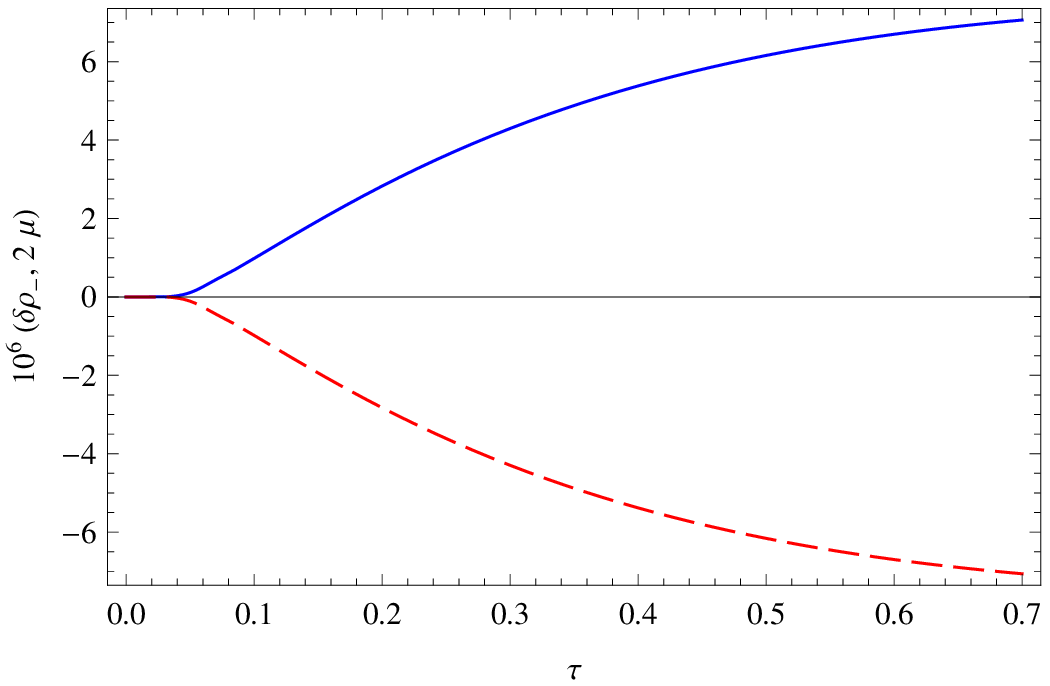} \\
(c)
\end{tabular}
\caption{Asymmetries in the sterile sector (a) (blue - solid line for 
$N_2$ and red - dashed line for $N_3$), in the active sector (b) (red 
- long line for $e$, blue - short line for $\mu$ and brown - dotted line for 
$\tau$) and total asymmetries in both sector (c) (blue - solid line for 
the sterile sector and red - dashed line for the active sector).}
\label{Asymmetries}
}

The asymmetries start to be produced around $\tau \simeq 0.05$ (see
Figs.~\ref{Asymmetries}a,b,c and below). The asymmetries in singlet
fermion $N_2$ is almost the same as in $N_3$
(Fig.~\ref{Asymmetries}a). The asymmetries in $\mu$ and $\tau$
flavours are very close to each other (upper line in Fig.
\ref{Asymmetries}b). They start to decrease for $\tau \gsim 0.5$,
which is the onset of damping (see below). The graph
(\ref{Asymmetries}c) shows us that  the asymmetry in the active sector
is the same with the opposite sign as  the one in the singlet sector.

In Fig. \ref{ImpE-t} we show the time evolution of the damping (left panel) and
$N_2-N_3$ oscillation factors (right panel), appearing as exponentials in
(\ref{change-of-variable}), for the relevant eigenvalue of the 
effective Hamiltonian $\mathcal{H}$: $\int \im(p_{E_2}) dt$ and  $\int
\re(p_{E_2}) dt$. The numerical and analytical results of the first expression
almost coincide for this particular choice of parameters, but not the real part
due to our choice of CP-violating phases.

The oscillation exponent Fig. \ref{ImpE-t} (right panel) reaches $1$ at $\tau
\simeq 0.05$, which is the time when the  asymmetries are produced on
the graphs of Figure~\ref{Asymmetries}. The damping exponent Fig.
\ref{ImpE-t} (left panel) approaches $1$ at $\tau \simeq 0.6$,  corresponding
roughly to maximum of the asymmetries in Fig.~\ref{Asymmetries}b.

In general, the results of numerical integration are in accordance
with the qualitative picture developed in \cite{Shaposhnikov:2008pf}. 

\FIGURE{
\centerline{
\includegraphics[width=0.45\textwidth]{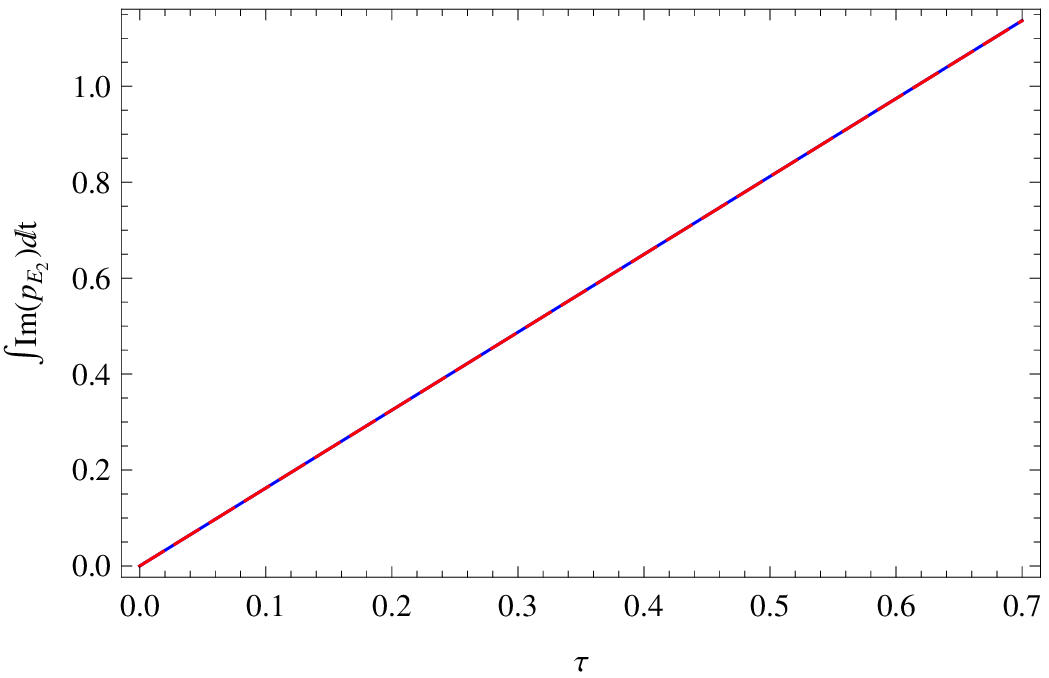}\ \ \ \
\includegraphics[width=0.45\textwidth]{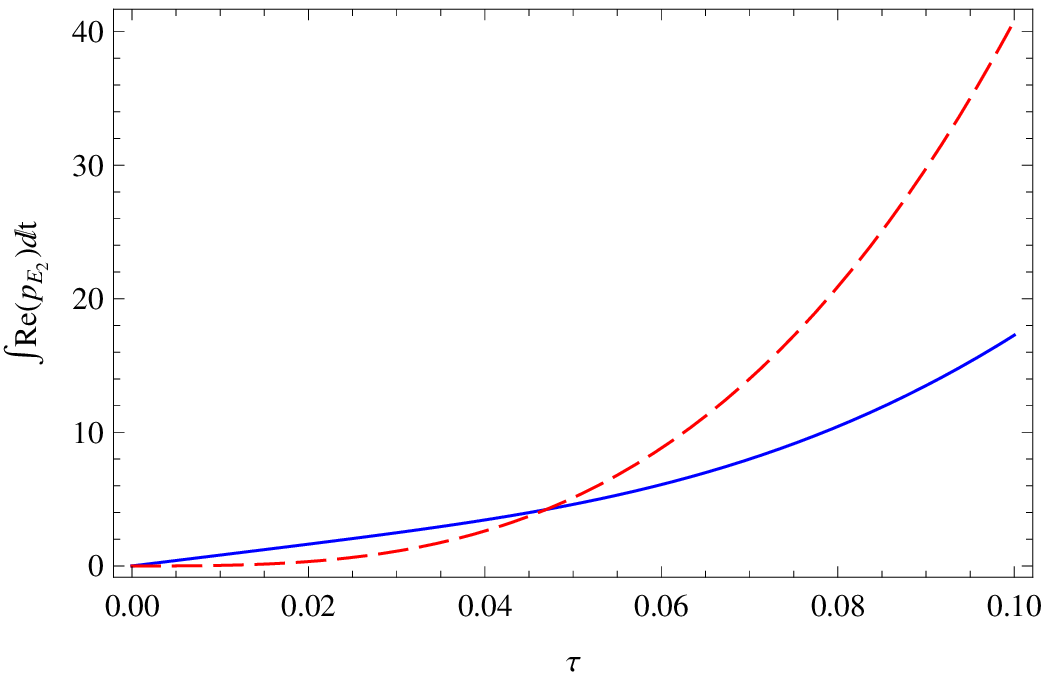}
}
\caption{Time dependence of $\int Im(p_{E_2}) dt$ (left) and of $\int
Re(p_{E_2}) dt$ (right). The numerical and analytical results are the
same (blue - solid line for the numerical results and red - dotted line 
for the analytical solution.}
\label{ImpE-t}
}
%
%
\section{CP phases and maximal baryon asymmetry}
\label{se:max}
Our procedure to find the parameter space of the \nuMSM \ leading to
acceptable baryon asymmetry of the Universe implies the maximisation
of the result with respect to three unknown phases $\eta,\phi$ and
$\alpha$ for normal hierarchy and $\eta,\phi$ and $\xi$ for
inverted. With the choice of active neutrino mass parameters 
(\ref{Active_neutrinos_parameters_1},\ref{Active_neutrinos_parameters_2})
the numerical analysis shows that it is achieved close to
$\eta=\phi=\alpha =\frac{\pi}{2}$ and $\xi=\frac{\pi}{4}$. (Note that for $\theta_{13}
= 0$ and  $\theta_{23} = \frac{\pi}{4}$ and parametrisation of active
neutrino mixing matrix as in \cite{Strumia:2006db} the Dirac phase
$\phi$ plays no role for inverted hierarchy.) 

In  Fig.~\ref{Phi-Alpha-Eta-normal} we show the dependence of
asymmetry on $\phi$ for a typical choice of $M,~\Delta M_M$, and
$\epsilon$ (left panel, fixing $\eta=\alpha=\frac{\pi}{2}$) and on
$\alpha$ (right panel, fixing $\eta=\phi =\frac{\pi}{2}$). In
Fig.~\ref{Phi-Alpha-Eta-inverted} we show the dependence of
asymmetry on  $\xi$ (fixing $\eta=\alpha =\frac{\pi}{2}$) for the
case of inverted hierarchy.
\FIGURE{
\centerline{
\includegraphics[width=0.45\textwidth]{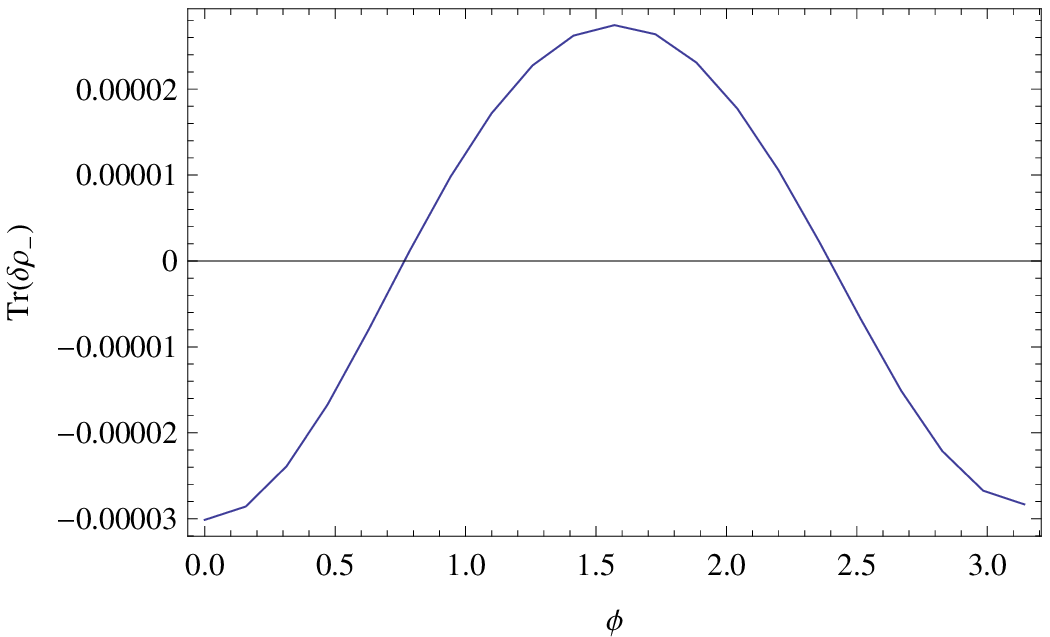}\ \ \ \
\includegraphics[width=0.45\textwidth]{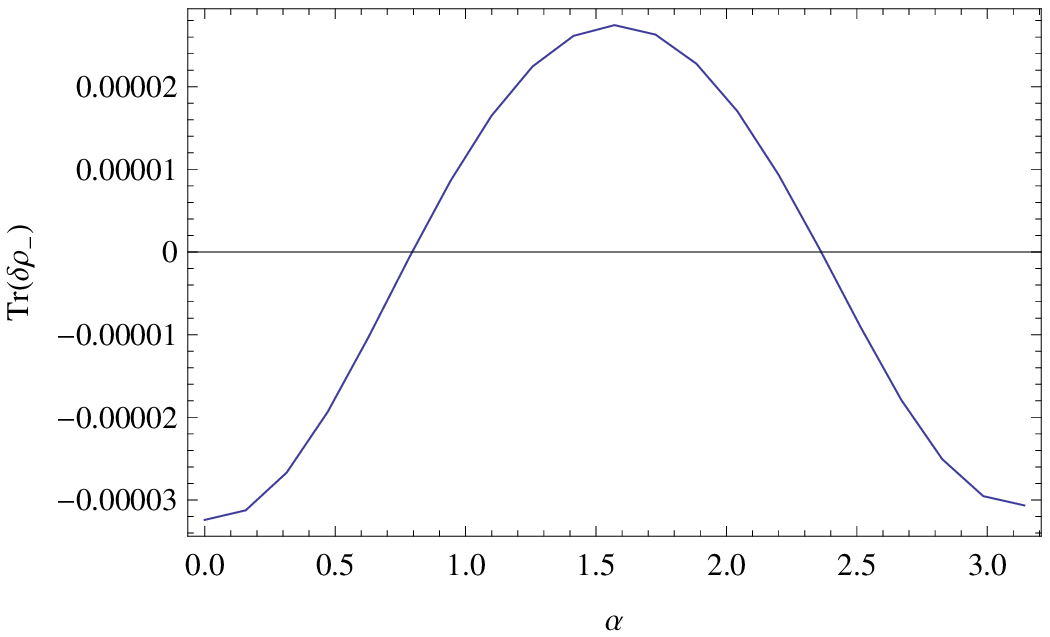}
}
\caption{Lepton asymmetry as a  function of $\phi$ (left panel) and $\alpha$
(right panel) for normal  hierarchy.}
\label{Phi-Alpha-Eta-normal}
}
\FIGURE{
\centerline{
\includegraphics[width=0.45\textwidth]{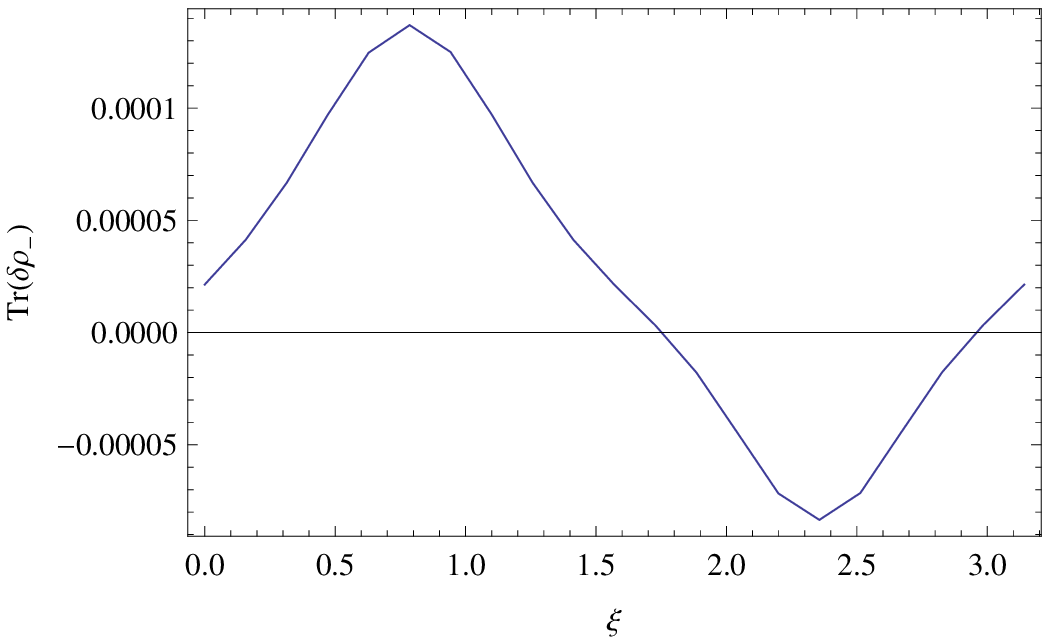}
}
\caption{Lepton asymmetry as a function of $\xi$ for inverted 
hierarchy.}
\label{Phi-Alpha-Eta-inverted}
}

The boundaries of the parameter space given in Sect.
\ref{se:baryon_asymmetry} correspond to the extremal choice of phases,
found in this Appendix and to the observed baryon asymmetry.  Below we
will determine the values of $M,~\Delta M_M$ and $\epsilon$ which
extremise asymmetry and find its maximum (using, again, the set of
phases (\ref{Phases})).

As a guiding line for numerics, we can proceed first with the
analytical estimate. In Fig. \ref{Max-asymmetry-deltaCP} we plot the
dependence of $\delta_{CP}$, defined in (\ref{deltaCP}) on $\epsilon$.
In general, CP-violation is larger in inverted hierarchy case and is
maximal around $\epsilon \simeq 0.5$. Replacing the brackets in 
(\ref{Asymmetry_approx}) by $1$, we arrive to
\be
\label{amax}
\frac{n_B}{s}\bigg|_{max}\simeq 1.5\times 10^{-3}~~(2.5\times 10^{-4})
\ee
for inverted (normal) hierarchy. The asymmetry is expected to be
maximal for
\be
\label{amax-exp}
\frac{\sin\varphi  \mathcal{F}^2 M_0}{8 \ \epsilon \ T_W}\sim 1,~~~
\frac{T^3_W}{4 M \Delta M_M M_0}\sim 1~.
\ee
\FIGURE{
\centerline{
\includegraphics[width=0.45\textwidth]{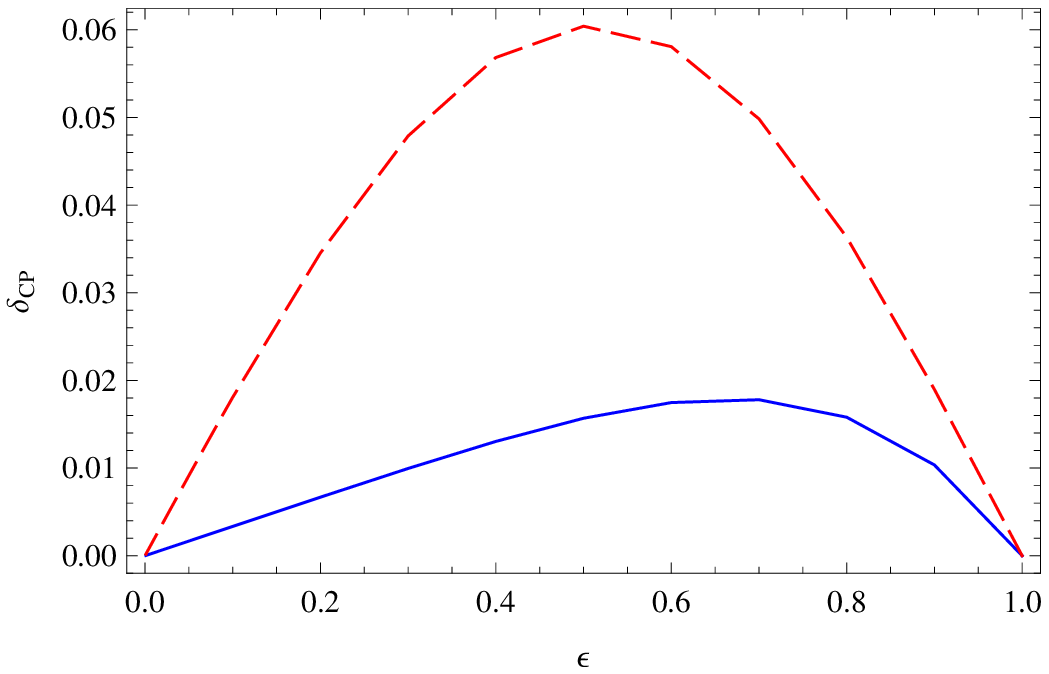}
}
\caption{$\delta_{CP}$ in function of $\epsilon$ for the normal hierarchy (blue 
- solid line) and for the inverted hierarchy (red - dashed line).}
\label{Max-asymmetry-deltaCP}
}

These expectations can be verified by numerics. In Fig.
\ref{Max-asymmetry-general} we show the dependence of the maximal
asymmetry on the singlet fermion mass (for each  choice of $M$ and
$T_W$ we find the values of $\epsilon$ and $\Delta M_M$ that give the
maximal asymmetry). 
\FIGURE{
\centerline{
\includegraphics[width=0.45\textwidth]{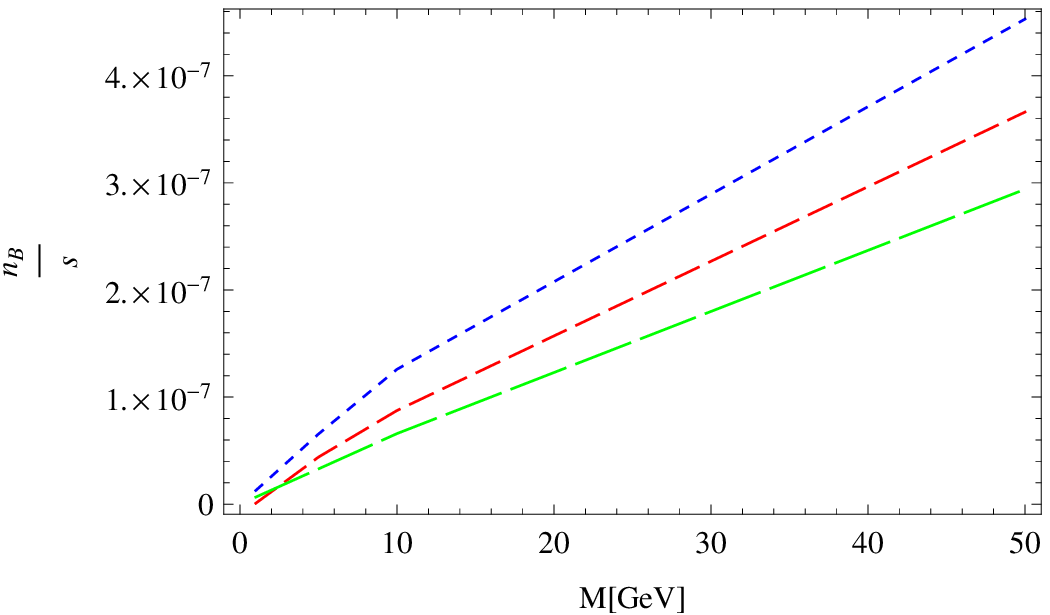}\ \ \ \
\includegraphics[width=0.45\textwidth]{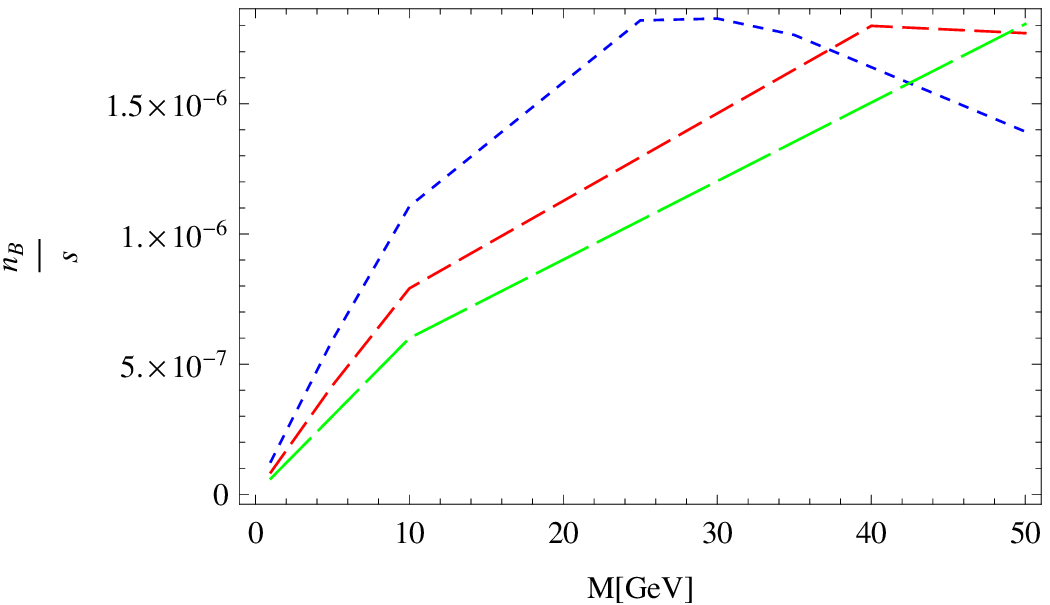}
}
\caption{Maximal asymmetry as a function of $M$ for $T_W = 100$ GeV (blue - 
short dashed line), $T_W = 150$ GeV (red - dashed line) and $T_W = 200$ GeV 
(green - long dashed line). Left panel -  normal hierarchy, right panel - inverted hierarchy.}
\label{Max-asymmetry-general}
}

In Fig. \ref{Max-asymmetry-epsilon-deltam} we show for which 
$\epsilon$ and $\Delta M_M$ the asymmetry is maximised for different
$T_W$, with the parameter along the curves being the mass $M$.
\FIGURE{
\centerline{
\includegraphics[width=0.45\textwidth]{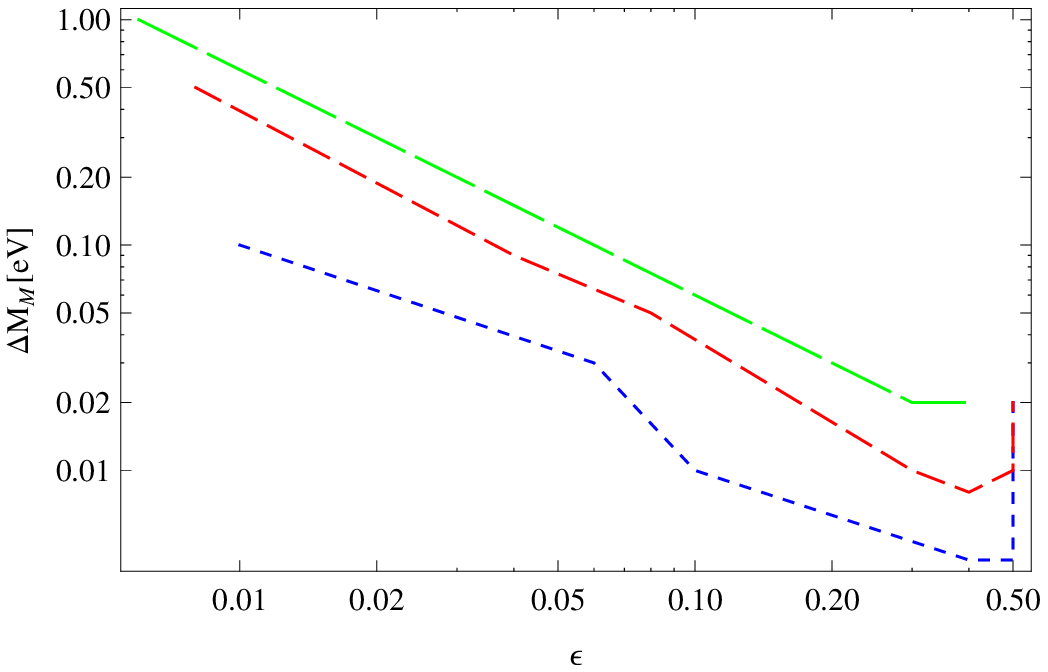}\ \ \ \
\includegraphics[width=0.45\textwidth]{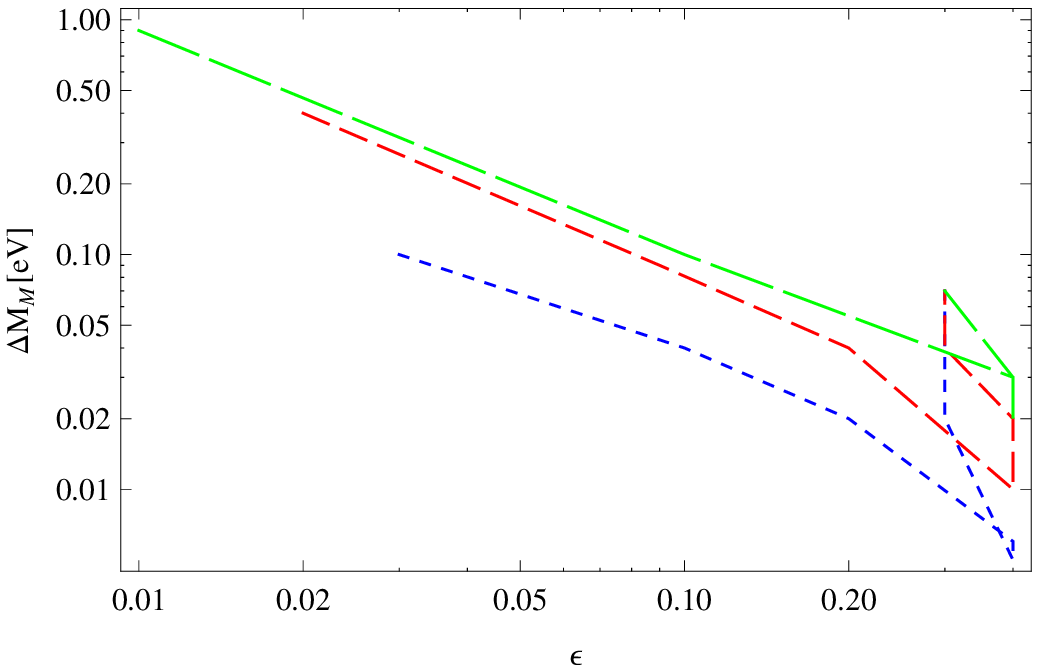}
}
\caption{The parameters $\epsilon$ and $\Delta M_M$ corresponding to 
the maximal asymmetry for $T_W = 100$ GeV (blue - short dashed line), 
$T_W = 150$ GeV (red - dashed line) and $T_W = 200$ GeV (green - 
long dashed line). The mass $M$ grows to the right Left panel -  normal
hierarchy, right panel - inverted hierarchy.}
\label{Max-asymmetry-epsilon-deltam}
}

In Fig. \ref{Max-asymmetry-general}, the maximal asymmetry for the
normal hierarchy is reached for masses $M$ bigger than $50$ GeV.
For the inverted hierarchy, the numerical computation of the asymmetry
maximum is $10^3$ times smaller than the estimate \eqref{amax}. This
difference comes from the fact that one can not extremise both
expressions in \eqref{amax-exp} at the same time for our choice of
temperature. Let us take the case $T = 100$ GeV for example. The
maximum (for the inverted hierarchy) is reached for $M = 25$ GeV,
$\epsilon = 0.5$ and $\Delta M_M = 5 \cdot10^{-12}$ GeV which
extremises the first expression of \eqref{amax-exp}, but the second
is of $O(10^{-3})$. Looking at the analytical expression, we should
decrease $\Delta M_M$ and get more asymmetry, but we are already
at the limit of validity of the perturbative expansion because of condition
\eqref{ana-exp-cond-3} of Section \ref{sse:lagr-ana-exp}.


\bibliography{bau_revised.bbl}

\end{document}